\documentclass[10pt,pra, twocolumn, nofootinbib, superscriptaddress, longbibliography]{revtex4-1}

\usepackage[T1]{fontenc}
\usepackage[latin9]{inputenc}
\usepackage{xcolor}
\usepackage{mathrsfs}
\usepackage{amsmath}
\usepackage{amssymb}
\usepackage{graphicx}
\usepackage{wasysym}
\usepackage[pdfusetitle,
 bookmarks=true,bookmarksnumbered=false,bookmarksopen=false,
 breaklinks=false,pdfborder={0 0 0},pdfborderstyle={},backref=false,colorlinks=true]
 {hyperref}
\hypersetup{
 citecolor=blue,urlcolor=blue}

\makeatletter

\newcommand{\noun}[1]{\textsc{#1}}

\usepackage{bbm}

\makeatother

\begin{document}
\title{An efficient framework for quantum dynamics driven by nonclassical
light }
\author{Sheng-Wen Li }
\email{lishengwen@bit.edu.cn }

\affiliation{Center for Quantum Technology Research, and Key Laboratory of Advanced
Optoelectronic Quantum Architecture and Measurements, School of Physics,
Beijing Institute of Technology, Beijing 100081, People\textquoteright s
Republic of China }
\author{Zeyang Liao}
\email{liaozy7@mail.sysu.edu.cn }

\affiliation{School of Physics, Sun Yat-sen University, Guangzhou 510275, China }
\author{Mao-Xin Liu}
\email{mxliu@bnu.edu.cn}

\affiliation{Beijing Normal University, Beijing 100081, People\textquoteright s
Republic of China }
\begin{abstract}
Understanding quantum system dynamics driven by nonclassical light
pulses is challenging, particularly for general light states with
large photon numbers. Here we introduce an efficient framework that
makes this task tractable. By introducing a pulse-shaped \emph{P}-representation,
the exact quantum evolution is decomposed into a mixture of many independent
quasi-classical branches, each governed by a standard master equation
with a classical pulse which can be solved efficiently. As an illustration,
for a two-level system interacting with an exponential pulse, we first
find out the exact analytical solutions to the Bloch equations in
each quasi-classical branch, and then by taking proper \emph{P}-function
average over all branches, the full system dynamics driven by nonclassical
light pulses is analytically obtained. For the one-photon and two-photon
cases, our method well reproduces the previous exact results either
analytically or numerically. Crucially, our approach scales efficiently
to more complex light states (Fock, thermal, squeezed vacuum states)
with large photon numbers ($N\sim100$). We further provide a clear
physical interpretation how the system dynamics is influenced through
the high-order optical coherence of the nonclassical pulses. This
work provides a unified and computationally efficient route and a
useful starting point to explore more complex quantum dynamics driven
by nonclassical light in quantum optics and quantum information processing. 
\end{abstract}
\maketitle

\section{Introduction }

Recent advances in the generation of nonclassical light have made
such quantum light sources increasingly accessible, spurring a growing
number of applications where the quantum nature of the input light
plays a central role. For instance, squeezed light has been shown
to enhance two-photon absorption by \textasciitilde\,47 times compared
to normal laser of the same intensity \citep{li_squeezed_2020}, and
to surpass the conventional cooling limit in laser cooling experiments
\citep{clark_sideband_2017}. Some reports show that the bright squeezed
vacuum light can be used to improve high-harmonic generation \citep{gorlach_high-harmonic_2023,rasputnyi_high-harmonic_2024}.
Distinct photon statistics have been shown to yield qualitatively
different signatures in scattering spectra \citep{vyas_resonance_1992}
and electron transport \citep{souquet_photon-assisted_2014}. These
examples indicate that the quantum nature of the driving light can
qualitatively affect the system behavior in ways beyond the classical
light \citep{schlawin_theory_2017,mukamel_roadmap_2020,raymer_entangled_2021,li_enhancement_2023}. 

Despite this progress, the theoretical methods to study the interaction
between a quantum system and a propagating multi-mode nonclassical
light pulse remain limited \citep{yang_non-classical_2021}. Usually,
the input light pulse is modeled as a classical wave packet, but such
a semi-classical treatment inherently discards the quantum features
such as photon statistics, high-order coherence, and cannot reflect
the distinction between classical and nonclassical light \citep{li_photon_2020}. 

Even for seemingly simple cases, such as a two-level system (TLS)
driven by a nonclassical Fock state pulse, how to efficiently calculate
the system dynamics still remains challenging \citep{mukhopadhyay_quantum_2024}.
Various formalisms have been developed to partially address this problem,
such as scattering matrix method \citep{Shen2005,Shen2007,Tsoi2008,Shen2015,Zheng2010,Fang2014},
wavefunction evolution approach \citep{Chen2011,Liao2015,Liao2016,Konyk2017,Liao2018,Dinc2019,Bundgaard2025},
Heisenberg operator approach \citep{Domokos2002,Barkemeyer2022,Regidor2025},
Green\textquoteright s function decomposition of the multi-particle
scattering matrix \citep{Laakso2014,Schneider2016}, input-output
formalism \citep{Fan2010,Lalumiere2013,Xu2015,Caneva2015,Kiilerich2019,zhou_dark-state-induced_2025},
Feynman diagrams \citep{Shi2009,Roulet2016,Piasotski2021,Dinc2020},
generalized master equation \citep{Gheri1998,Shi2015,You2018,Liao2020,yang_quantum_2024},
and tensor network method \citep{Pichler2016,Guimond2017,Regidor2021}.
Since most of these methods involves high order expansions with photon
numbers, solving the system dynamics becomes extremely challenging
when the photon number is very large (generally $N\apprle10$). 

In this work, we address this issue by introducing a systematic framework
established from the fully quantized EM field \citep{yao_enhancing_2020,cao_quantum_2022}.
By introducing a ``pulse-shaped'' \emph{P}-function for the nonclassical
input light \citep{sudarshan_equivalence_1963,glauber_coherent_1963,loudon_quantum_2000},
the full quantum evolution is decomposed into a mixture of many independent
quasi-classical ``branches''. Each branch describes the system's
response to a coherent wave packet as the input pulse, which can be
solved efficiently. The full quantum dynamics and observable expectations
are then recovered through the \emph{P}-function average over all
the evolution branches. Since the \emph{P}-functions of nonclassical
light either contain negative parts, or are highly singular generalized
functions \citep{schleich_quantum_2001,vogel_quantum_2006,agarwal_quantum_2012},
that makes the system dynamics different from that driven by classical
light \citep{yao_enhancing_2020,cao_quantum_2022}. 

To verify the reliability of this approach, we study the TLS dynamics
interacting with a nonclassical pulse. When interacting with a coherent
wave packet, we find an analytical solution for the TLS evolution;
then the dynamics driven by nonclassical light is obtained by proper
\emph{P}-function averaging. For the one-photon \citep{Shen2005,stobinska_perfect_2009,wang_efficient_2011}
and two-photon pulses \citep{Gheri1998,Shi2015,You2018,Liao2020,yang_quantum_2024},
our results show perfect agreement with known exact analytical and
numerical solutions, which well confirms the validity of our approach. 

Now we can efficiently obtain the TLS dynamics driven by a Fock pulse
with a quite large photon number ($N\sim100$ in our demonstration),
as well as thermal, squeezed vacuum pulses. The \emph{P}-function
properties also help us find a clear physical picture how the system
dynamics is influenced via the high order coherence of the input light
\citep{yao_enhancing_2020,cao_quantum_2022}.

The paper is organized as follows. In Sec.\,\ref{sec:General}, we
introduce the general framework about how to study the interaction
with a generic light state with the help of the pulse-shaped \emph{P}-function
and the corresponding decomposition. In Sec.\,\ref{sec:class}, we
derive the analytical quasi-classical solutions for a two-level system
interacting with a coherent pulse. In Sec.\,\ref{sec:nonclassical},
we apply the \emph{P}-function averaging method to various situations
using nonclassical light states as the input pulse, including the
Fock states and squeezed vacuum states. Finally the summary is given
in Sec.\,\ref{sec:conclusion}.

\section{General framework: interacting with a generic light pulse \label{sec:General}}

Here we give a general theoretical framework about how to study the
dynamical behavior of a multi-level quantum system interacting with
a generic light pulse, especially the nonclassical light states. To
deal with a general nonclassical light state, the EM field must be
described by the fully quantized theory as our starting point. Besides,
under proper conditions, such a general framework should return the
quasi-classical driving theory widely adopted before \citep{yao_enhancing_2020,cao_quantum_2022}. 

Generally, the Hamiltonian of the quantized EM field is $\hat{H}_{\text{\textsc{e}}}=\sum_{\mathbf{k}\varsigma}\,\hbar\omega_{\mathbf{k}}\,\hat{a}^{\dag}_{\mathbf{k}\varsigma}\hat{a}_{\mathbf{k}\varsigma}$,
and the interaction between a multi-level system and the EM field
is described by\footnote{In this paper, we use $\hat{o}$, $\hat{o}(t)$
and $\tilde{o}(t)$ to denote operators in the Schr\"odinger, Heisenberg
and interaction pictures respectively.} \citep{yao_enhancing_2020,cao_quantum_2022}
\begin{equation}
\begin{split}\tilde{V}_{\text{\textsc{se}}} & (t)=-\tilde{\mathbf{d}}(t)\cdot\tilde{\mathbf{E}}(\mathbf{r}_{0},t)\\
= & -\tilde{\mathbf{d}}(t)\cdot\sum_{\mathbf{k},\varsigma}\vec{\mathrm{e}}_{\mathbf{k}\varsigma}\sqrt{\frac{\hbar\omega_{\mathbf{k}}}{2\epsilon_{0}V}}\,\big(i\,\hat{a}_{\mathbf{k}\varsigma}\,e^{i\mathbf{k}\cdot\mathbf{r}_{0}-i\omega_{\mathbf{k}}t}+\mathrm{H.c.}\big).
\end{split}
\label{eq:V_SE}
\end{equation}
Here, $\tilde{\mathbf{d}}(t)$ is the electric dipole operator of
the multilevel system, with $\mathbf{r}_{0}$ as its position; $\tilde{\mathbf{E}}(\mathbf{r},t)$
is the quantized electric field operator, $\varsigma$ is the polarization
index, and $\hat{a}_{\mathbf{k}\varsigma}$ is the annihilation operator
for the field mode-$(\mathbf{k}\varsigma)$ {[}hereafter we use $\mathtt{k}$
to represent the mode index $(\mathbf{k}\varsigma)$ for simplicity{]}. 

\subsection{Quasi-classical driving interaction \label{subsec:class-driving}}

We first consider a particular situation that the initial quantum
state of the EM field is a multi-mode coherent pulse state, that is
\citep{loudon_quantum_2000,cao_quantum_2022}, 
\begin{equation}
\begin{split}\big|\alpha,\,\{\beta_{\mathtt{k}}\}\big\rangle:= & \prod_{\mathtt{k}}\hat{D}_{\mathtt{k}}(\beta_{\mathtt{k}}\,\alpha)\,|\varnothing\rangle\\
= & \dots\otimes|\beta_{\mathtt{k}}\cdot\alpha\rangle_{\mathtt{k}}\otimes|\beta_{\mathtt{k}'}\cdot\alpha\rangle_{\mathtt{k}'}\otimes\dots
\end{split}
\label{eq:coherent-pulse}
\end{equation}
Here $|\varnothing\rangle$ is the vacuum state of the quantized EM
field, $\hat{D}_{\mathtt{k}}(\lambda):=\exp(\lambda\,\hat{a}^{\dag}_{\mathtt{k}}-\lambda^{*}\,\hat{a}_{\mathtt{k}})$
is the displacement operator for the field mode-$\mathtt{k}$, and
the set $\{\beta_{\mathtt{k}}\}$ is normalized as $\sum_{\mathtt{k}}\,|\beta_{\mathtt{k}}|^{2}=1$. 

In this state, each field mode-$\mathtt{k}$ stays in a coherent state
$|\beta_{\mathtt{k}}\,\alpha\rangle_{\mathtt{k}}$, where $\{\beta_{\mathtt{k}}\}$
describes the ``shape'' profile in the frequency domain, while $\alpha$
describes its total amplitude. The total photon number is $\big\langle\sum\hat{a}^{\dag}_{\mathtt{k}}\hat{a}_{\mathtt{k}}\big\rangle_{\alpha,\{\beta_{\mathsf{k}}\}}=\sum_{\mathtt{k}}\,|\beta_{\mathtt{k}}\,\alpha|^{2}=|\alpha|^{2}$.
When the EM field is in such a coherent pulse state (\ref{eq:coherent-pulse}),
the mean value of the electric field operator gives 
\begin{align}
\langle\hat{\mathbf{E}}(\mathbf{r},t)\rangle_{\alpha,\{\beta_{\mathtt{k}}\}} & =\sum_{\mathbf{k},\varsigma}\vec{\mathrm{e}}_{\mathbf{k}\varsigma}\sqrt{\frac{\hbar\omega_{\mathbf{k}}}{2\epsilon_{0}V}}\,\big(i\,\alpha\beta_{\mathbf{k}\varsigma}\,e^{i\mathbf{k}\cdot\mathbf{r}-i\omega_{\mathbf{k}}t}+\mathrm{c.c.}\big)\nonumber \\
 & :=\vec{E}_{\alpha}(\mathbf{r},t),\label{eq:E(rt)}
\end{align}
which corresponds to a classical wave packet. 

On the other hand, the field operator $\hat{a}_{\mathtt{k}}$ can
be divided as its mean value $\langle\hat{a}_{\mathtt{k}}\rangle_{\alpha,\{\beta_{\mathtt{k}}\}}=\beta_{\mathtt{k}}\,\alpha$
and the pure quantum fluctuation $\delta\hat{a}_{\mathtt{k}}:=\hat{a}_{\mathtt{k}}-\langle\hat{a}_{\mathtt{k}}\rangle_{\alpha,\{\beta_{\mathtt{k}}\}}$.
It can be verified that $\langle\delta\hat{a}^{\dag}_{\mathtt{q}}\,\delta\hat{a}^{(\dag)}_{\mathtt{q}'}\rangle_{\alpha,\{\beta_{\mathtt{k}}\}}=\langle\varnothing|\hat{a}^{\dag}_{\mathtt{q}}\,\hat{a}^{(\dag)}_{\mathtt{q}'}|\varnothing\rangle$,
which means the fluctuations properties of $\delta\hat{a}_{\mathtt{q}}$
in the pulse state $|\alpha,\{\beta_{k}\}\rangle$ are the same as
the field operator $\hat{a}_{\mathtt{q}}$ in the vacuum state $|\varnothing\rangle$.

Correspondingly, the electric field operator also can be divided as
$\tilde{\mathbf{E}}(\mathbf{r},t)\equiv\vec{E}_{\alpha}(\mathbf{r},t)+\tilde{\mathbf{E}}^{(0)}(\mathbf{r},t)$,
and the above system-field\noun{ }interaction (\ref{eq:V_SE}) can
be rewritten as $\tilde{V}_{\text{\textsc{se}}}=\tilde{V}_{\alpha}(t)+\tilde{V}^{(0)}_{\text{\textsc{se}}}(t)$,
where $\tilde{V}_{\alpha}(t)=-\tilde{\mathbf{d}}(t)\cdot\vec{E}_{\alpha}(\mathbf{r}_{0},t)$
and \citep{yao_enhancing_2020,cao_quantum_2022}
\begin{align}
\tilde{V}^{(0)}_{\text{\textsc{se}}} & (t)=-\tilde{\mathbf{d}}(t)\cdot\tilde{\mathbf{E}}^{(0)}(\mathbf{r}_{0},t)\label{eq:V-alpha}\\
= & -\tilde{\mathbf{d}}(t)\cdot\sum_{\mathbf{k},\varsigma}\vec{\mathrm{e}}_{\mathbf{k}\varsigma}\sqrt{\frac{\hbar\omega_{\mathbf{k}}}{2\epsilon_{0}V}}\,\big(i\,\delta\hat{a}_{\mathbf{k}\varsigma}\,e^{i\mathbf{k}\cdot\mathbf{r}_{0}-i\omega_{\mathbf{k}}t}+\mathrm{H.c.}\big).\nonumber 
\end{align}
Clearly, $\tilde{V}_{\alpha}(t)$ indicates the quasi-classical interaction
between the electric dipole and a wave packet $\vec{E}_{\alpha}(\mathbf{r}_{0},t)$,
while $\tilde{V}^{(0)}_{\text{\textsc{se}}}(t)$ describes the contributions
from the pure quantum fluctuations of the EM field. 

To study the dynamics of the multi-level system $\hat{\rho}_{\text{\textsc{s}}}(t)=\mathrm{tr}_{\text{\textsc{e}}}\big[\hat{\boldsymbol{\rho}}_{\text{\textsc{se}}}(t)\big]$,
we can take proper integral iteration of the von Neumann equation
$\partial_{t}\tilde{\boldsymbol{\rho}}_{\text{\textsc{se}}}=\frac{i}{\hbar}[\tilde{\boldsymbol{\rho}}_{\text{\textsc{se}}},\,\tilde{V}_{\alpha}+\tilde{V}^{(0)}_{\text{\textsc{se}}}]$,
and apply the Born-Markov approximations \citep{breuer_theory_2002,li_non-markovianity_2016}.
Since the fluctuation properties of $\tilde{\mathbf{E}}^{(0)}(\mathbf{r},t)$
in $\tilde{V}^{(0)}_{\text{\textsc{se}}}(t)$ are the same with the
field operator when studying the the spontaneous emission without
any driving light, these treatments finally lead to the following
master equation (see Appendix \ref{sec:Master-equation-derivation}
and Refs.\,\citep{yao_enhancing_2020,cao_quantum_2022}) 
\begin{equation}
\partial_{t}\tilde{\mathbf{\rho}}^{(\alpha)}_{\text{\textsc{s}}}(t)=\frac{i}{\hbar}[\,\tilde{\mathbf{\rho}}^{(\alpha)}_{\text{\textsc{s}}}(t),\,\tilde{V}_{\alpha}(t)\,]+\mathcal{L}[\,\tilde{\mathbf{\rho}}^{(\alpha)}_{\text{\textsc{s}}}(t)\,].\label{eq:ME-alph}
\end{equation}
 Here we add a superscript ``$(\alpha)$'' to the system state so
as to emphasize such a dynamics is obtained when the EM field starts
from the coherent pulse state $|\alpha,\{\beta_{k}\}\rangle$. The
first term describes the quasi-classical driving interaction with
the wave packet, and $\mathcal{L}[\,\rho\,]$ is the same with the
standard GKSL (Lindblad) dissipation terms when studying the spontaneous
emission without any driving light \citep{gorini_completely_1976,lindblad_generators_1976},
which does not depend on $\alpha$ or $\{\beta_{\mathtt{k}}\}$. 

The above equation (\ref{eq:ME-alph}) is just the master equation
with the quasi-classical driving interaction widely adopted in literature.
In this sense, when the initial state of the EM field is the above
multi-mode coherent pulse state, our fully quantum treatment well
returns the previous quasi-classical approach. 

\subsection{Interacting with a nonclassical light pulse \label{subsec:nonclassical}}

Now we study the more general situation that the input light pulse
is not the above coherent state (\ref{eq:coherent-pulse}). In this
case, the above master equation (\ref{eq:ME-alph}) for quasi-classical
driving is clearly no longer sufficient enough to enclose the specific
statistical properties of the driving light, especially the high order
nonclassical features.

First, we introduce a ``pulse-shaped'' \emph{P}-function to represent
the quantum state of a generic nonclassical light pulse, 
\begin{align}
\hat{\boldsymbol{\rho}}_{\mathnormal{\textsc{e}}} & =\int\text{d}^{2}\alpha\,P(\alpha,\alpha^{*})\,\big|\alpha,\,\{\beta_{\mathtt{k}}\}\big\rangle\big\langle\alpha,\,\{\beta_{\mathtt{k}}\}\big|\nonumber \\
 & =\int\text{d}^{2}\alpha\,P(\alpha,\alpha^{*})\,\hat{\boldsymbol{\rho}}^{(\alpha)}_{\text{\textsc{e}}}.\label{eq:P-pulse}
\end{align}
 Here, $\hat{\boldsymbol{\rho}}^{(\alpha)}_{\text{\textsc{e}}}\equiv\big|\alpha,\,\{\beta_{\mathtt{k}}\}\big\rangle\big\langle\alpha,\,\{\beta_{\mathtt{k}}\}\big|$
is a coherent pulse state (\ref{eq:coherent-pulse}), with $\alpha$
as the total amplitude and $\{\beta_{\mathtt{k}}\}$ as its shape,
which corresponds to a classical wave packet $\vec{E}_{\alpha}(\mathbf{r},t)$
{[}Eq.\,(\ref{eq:E(rt)}){]}. Formally, the full density state $\hat{\boldsymbol{\rho}}_{\mathnormal{\textsc{e}}}$
could be regarded as the mixture of many coherent pulses $\hat{\boldsymbol{\rho}}^{(\alpha)}_{\text{\textsc{e}}}$,
with $P(\alpha,\alpha^{*})$ as the ``quasi-probability'' \citep{agarwal_quantum_2012,vogel_quantum_2006,walls_quantum_2008}.
Thus, $\hat{\boldsymbol{\rho}}_{\mathnormal{\textsc{e}}}$ also has
the same pulse shape determined by $\{\beta_{\mathtt{k}}\}$. 

Besides, $P(\alpha,\alpha^{*})$ is the Glauber-Sudarshan \emph{P}-function
widely used in quantum optics \citep{sudarshan_equivalence_1963,glauber_coherent_1963},
and here it only depends on the total amplitude $\alpha$ but not
on the pulse shape $\{\beta_{\mathtt{k}}\}$. When $P(\alpha,\alpha^{*})$
is a positive-definite function, $\hat{\boldsymbol{\rho}}_{\mathnormal{\textsc{e}}}$
is called a ``classical'' light state (e.g., the thermal state);
if $P(\alpha,\alpha^{*})$ contains negative parts, or is more singular
than $\delta$-function, $\hat{\boldsymbol{\rho}}_{\mathnormal{\textsc{e}}}$
is called a ``nonclassical'' light state (e.g., the Fock states,
see Appendix \ref{sec:Apx-pulse}) \citep{vogel_quantum_2006,walls_quantum_2008,agarwal_quantum_2012}.

In such a representation (\ref{eq:P-pulse}) for a pulse-shaped \emph{P}-function,
the critical advantage is, the multi-mode pulse shape is fully described
by $\{\beta_{\mathtt{k}}\}$, while the nonclassical quantum features
of the light state are enclosed well in $P(\alpha,\alpha^{*})$, which
only depends on the total amplitude $\alpha$ but not the shape. 

If the initial state of the EM field is such a generic pulse state
(\ref{eq:P-pulse}), the system dynamics $\hat{\rho}_{\text{\textsc{s}}}(t)=\mathrm{tr}_{\text{\textsc{e}}}\big[\hat{\boldsymbol{\rho}}_{\text{\textsc{se}}}(t)\big]$
can be obtained from the unitary evolution as \citep{yao_enhancing_2020,cao_quantum_2022},
\begin{align}
\hat{\rho}_{\text{\textsc{s}}}(t) & =\mathrm{tr}_{\text{\textsc{e}}}\big[\,\mathcal{U}_{t}\,\hat{\rho}_{\text{\textsc{s}}}(0)\otimes\hat{\boldsymbol{\rho}}_{\text{\textsc{e}}}(0)\,\mathcal{U}^{\dagger}_{t}\,\big]\nonumber \\
 & =\int\text{d}^{2}\alpha\,P(\alpha,\alpha^{*})\,\mathrm{tr}_{\text{\textsc{e}}}\big[\,\mathcal{U}_{t}\,\hat{\rho}_{\text{\textsc{s}}}(0)\otimes\hat{\boldsymbol{\rho}}^{(\alpha)}_{\mathnormal{\textsc{e}}}\,\mathcal{U}^{\dagger}_{t}\,\big]\nonumber \\
 & :=\int\text{d}^{2}\alpha\,P(\alpha,\alpha^{*})\,\hat{\rho}^{(\alpha)}_{\text{\textsc{s}}}(t),\label{eq:unitary}
\end{align}
Here, $\hat{\rho}^{(\alpha)}_{\text{\textsc{s}}}(t):=\mathrm{tr}_{\text{\textsc{e}}}\big[\,\mathcal{U}_{t}\,\hat{\rho}_{\text{\textsc{s}}}(0)\otimes\hat{\boldsymbol{\rho}}^{(\alpha)}_{\mathnormal{\textsc{e}}}\,\mathcal{U}^{\dagger}_{t}\,\big]$
indicates the system evolution if the EM field starts from a specific
coherent pulse state $\hat{\boldsymbol{\rho}}^{(\alpha)}_{\mathnormal{\textsc{e}}}$.
It is worth noting that indeed such a dynamical process just can be
described by the master equation (\ref{eq:ME-alph}), where $\alpha$
gives the driving amplitude in $\tilde{V}_{\alpha}(t)$. 

Thus, Eq.\,(\ref{eq:unitary}) indicates that the full dynamics for
$\hat{\rho}_{\text{\textsc{s}}}(t)$ can be obtained as the \emph{P}-function
average of many evolution ``\emph{branches}'', and we call $\hat{\rho}^{(\alpha)}_{\text{\textsc{s}}}(t)$
as the \emph{quasi-classical $\alpha$-branch} of the full evolution
\citep{yao_enhancing_2020,cao_quantum_2022}. The dynamics of each
$\alpha$-branch just can be described by the master equation (\ref{eq:ME-alph})
with the quasi-classical driving interaction. We emphasize that the
the above decomposition (\ref{eq:unitary}) is exact, though generally
it is not easy to obtain the dynamics $\hat{\rho}^{(\alpha)}_{\text{\textsc{s}}}(t)$
for each $\alpha$-branch exactly. In principle, approximated or numerical
solutions could also give well enough estimations. 

Similarly, the observable expectations of the system also can be obtained
as 
\begin{align}
\langle\hat{O}_{\text{\textsc{s}}}(t)\rangle & =\int\text{d}^{2}\alpha\,P(\alpha,\alpha^{*})\,\langle\hat{O}^{(\alpha)}_{\text{\textsc{s}}}(t)\rangle,\label{eq:O-average}
\end{align}
 where $\langle\hat{O}^{(\alpha)}_{\text{\textsc{s}}}(t)\rangle:=\mathrm{tr}_{\text{\textsc{s}}}\big[\,\hat{O}_{\text{\textsc{s}}}\cdot\hat{\rho}^{(\alpha)}_{\text{\textsc{s}}}(t)\,\big]$
can be calculated by the master equation (\ref{eq:ME-alph}) for the
$\alpha$-branch $\hat{\rho}^{(\alpha)}_{\text{\textsc{s}}}(t)$. 

In sum, when the initial state of the EM field is a generic nonclassical
pulse, the algorithm is: we first decompose the input light state
as the above \emph{P}-function mixture (\ref{eq:P-pulse}), then we
solve the quasi-classical master equation (\ref{eq:ME-alph}) for
each $\alpha$-branch $\hat{\rho}^{(\alpha)}_{\text{\textsc{s}}}(t)$
(either analytically or numerically), and finally their \emph{P}-function
average gives the full system dynamics $\hat{\rho}_{\text{\textsc{s}}}(t)$
driven by the nonclassical light pulse \citep{yao_enhancing_2020,cao_quantum_2022}.

\begin{figure}
\includegraphics[width=0.7\columnwidth]{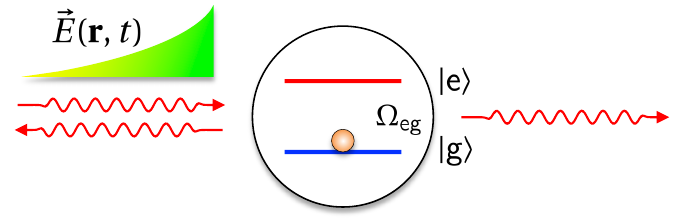}

\caption{Demonstration for a TLS interacting with a classical light pulse described
by a wave packet $\vec{E}(\mathbf{r},t)$.}
\label{fig-TLS-pulse}
\end{figure}

\section{A two-level system driven by a generic light pulse \label{sec:class}}

Based on the above general formalism, now we study a specific problem,
that is, the dynamics of a TLS scattered by a generic light pulse,
which could be either a classical coherent wave packet, or a nonclassical
light state, such as the single photon or multi-photon Fock state.
For simplicity, hereafter we consider the EM field is 1-dimensional
with only one polarization direction $\vec{\mathrm{e}}_{\text{p}}$. 

\subsection{Basic setup for the quasi-classical driving }

First, we study the quasi-classical driving situation (Sec.\,\ref{subsec:class-driving}).
We assume the initial state of the 1D EM field is a coherent pulse
state $|\alpha,\{\beta_{k}\}\rangle$, and the pulse shape $\{\beta_{k}\}$
take the following Lorentzian form in the frequency domain, 
\begin{eqnarray}
\beta_{k} & = & \sqrt{\frac{c\kappa}{L}}\frac{1}{ck-\omega_{\text{d}}+i\kappa/2},\label{eq:beta-k}\\
\sum_{k}|\beta_{k}|^{2} & \stackrel{L\rightarrow\infty}{\longrightarrow} & \frac{L}{2\pi}\int^{\infty}_{-\infty}\text{d}k\,\frac{c\kappa/L}{(ck-\omega_{\text{d}})^{2}+\kappa^{2}/4}=1.\nonumber 
\end{eqnarray}
Here $\omega_{\text{d}}$ is the central frequency of the light pulse,
$\kappa$ is the linewidth ($\omega_{\text{d}}\gg\kappa$), $k$ is
the wave vector of the normal mode ($\omega_{k}=c|k|$), and $L$
is the length of the 1D EM field. 

For this coherent pulse state, such a distribution $\{\beta_{k}\}$
gives an exponentially decaying pulse shape in the time domain. For
the electric field (\ref{eq:E(rt)}), its expectation $\vec{E}_{\alpha}(x,t)\equiv\frac{1}{2}\big[\vec{E}^{(+)}_{\alpha}(x,t)+\vec{E}^{(-)}_{\alpha}(x,t)\big]$
gives 
\begin{align}
 & \vec{E}^{(+)}_{\alpha}(x,t):=\sum_{k}\vec{\mathrm{e}}_{\text{p}}\sqrt{\frac{\hbar\omega_{k}}{2\epsilon_{0}V}}\,i\,\alpha\beta_{k}\,e^{ikx-i\omega_{k}t}\label{eq:E(xt)}\\
 & \stackrel{L\rightarrow\infty}{\longrightarrow}i\alpha\,\vec{\mathrm{e}}_{\text{p}}\cdot\frac{L}{2\pi}\int\text{d}k\,\sqrt{\frac{2\hbar\omega_{k}}{\epsilon_{0}V}}\,\frac{\sqrt{c\kappa/L}}{ck-\omega_{\text{d}}+i\kappa/2}\,e^{-ik(ct-x)}\nonumber \\
 & =\vec{\mathrm{e}}_{\text{p}}\,\alpha\mathtt{E}_{0}\,\Theta(t-\frac{x}{c})\,e^{(-i\omega_{\text{d}}-\frac{1}{2}\kappa)(t-\frac{x}{c})},\quad\mathtt{E}_{0}:=\sqrt{\frac{2\hbar\omega_{\text{d}}}{\epsilon_{0}(\mathsf{S}c/\kappa)}}.\nonumber 
\end{align}
Here $\vec{E}^{(\pm)}_{\alpha}(x,t)$ are complex conjugations of
each other. $\Theta(t)$ is the Heaviside function which gives the
wave front. Clearly, $\vec{E}_{\alpha}(x,t)$ has a propagating form
with an exponentially decaying shape, with $\kappa^{-1}$ as the pulse
length in the time domain (see demonstration in Fig.\,\ref{fig-TLS-pulse}). 

Besides, in the above wave packet $\vec{E}_{\alpha}(x,t)$, $\mathtt{E}_{0}$
can be regarded as the effective field amplitude generated by a single
photon. $V$ is the volume of the whole EM field, and $\mathsf{S}\equiv V/L$
can be regarded as the effective cross section of this 1D EM field.
Therefore, the quantity $(\mathsf{S}c/\kappa)$ in $\mathtt{E}_{0}$
can be effectively regarded as the ``volume'' of this light pulse. 

For a TLS staying at $x=0$ with the dipole operator $\tilde{\mathbf{d}}(t)=\vec{\wp}\,\tilde{\sigma}^{x}(t)$,
based on the discussions around Eq.\,(\ref{eq:V-alpha}), its interaction
with the wave packet $\vec{E}_{\alpha}(x,t)$ is
\begin{align}
\tilde{V}_{\alpha}(t) & =-\vec{\wp}\,\tilde{\sigma}^{x}(t)\cdot\vec{E}_{\alpha}(t)\nonumber \\
 & \simeq-\frac{1}{2}\big[\hbar\mathcal{E}_{\alpha}\,\hat{\sigma}^{+}e^{i(\Omega_{\mathsf{eg}}-\omega_{\text{d}})t}+\mathrm{H.c.}\big]\,e^{-\frac{1}{2}\kappa t},\label{eq:V_a-TLS}\\
\hbar\mathcal{E}_{\alpha} & :=(\vec{\wp}\cdot\vec{\mathrm{e}}_{\text{p}})\,\mathtt{E}_{0}\,\alpha\equiv\alpha\,\wp\sqrt{\frac{2\hbar\omega_{\text{d}}}{\epsilon_{0}(\mathsf{S}c/\kappa)}}\equiv\alpha\,\eta_{0}.\nonumber 
\end{align}
Here we denote $\wp\equiv\vec{\wp}\cdot\vec{\mathrm{e}}_{\text{p}}$
for the dipole moment (assuming $\wp>0$ without loss of generality).
$\hbar\Omega_{\mathsf{eg}}$ is the energy gap between the two energy
levels $|\mathsf{e/g}\rangle$. $\hbar\mathcal{E}_{\alpha}\equiv\alpha\,\eta_{0}$
is the coupling strength between the TLS and the light pulse, where
$\eta_{0}\equiv\wp\,\mathtt{E}_{0}$ can be regarded as the effective
coupling strength with a single photon. In the second line, the rotating-wave
approximation (RWA) is applied. 

With these preparations, based on the above discussions around Eqs.\,(\ref{eq:V-alpha},
\ref{eq:ME-alph}), the dynamical behavior $\tilde{\rho}^{(\alpha)}_{\text{\textsc{s}}}(t)$
of the TLS in this specific ``$\alpha$-branch'' is described by
the master equation (see Appendix \ref{sec:Master-equation-derivation}
and Refs.\,\citep{yao_enhancing_2020,cao_quantum_2022}) 
\begin{align}
\partial_{t}\tilde{\rho}^{(\alpha)}_{\text{\textsc{s}}} & =\frac{i}{\hbar}[\tilde{\rho}^{(\alpha)}_{\text{\textsc{s}}},\,\tilde{V}_{\alpha}(t)]+\mathcal{L}[\tilde{\rho}^{(\alpha)}_{\text{\textsc{s}}}],\nonumber \\
\mathcal{L}[\rho] & =\gamma\big(\hat{\sigma}^{-}\rho\hat{\sigma}^{+}-\frac{1}{2}\hat{\sigma}^{+}\hat{\sigma}^{-}\rho-\frac{1}{2}\rho\hat{\sigma}^{+}\hat{\sigma}^{-}\big).\label{eq:TLS-ME}
\end{align}
Here the superscript ``$\alpha$'' is remained, indicating this
equation describes the situation that the EM field starts from the
coherent pulse $|\alpha,\{\beta_{k}\}\rangle$. $\tilde{V}_{\alpha}(t)$
is the quasi-classical driving interaction (\ref{eq:V_a-TLS}), which
depends on $\alpha$. The dissipation term $\mathcal{L}[\rho]$ is
the same with the standard GKSL (Lindblad) term when studying the
spontaneous emission of the TLS in absence of any light driving, and
the super-operator $\mathcal{L}[\:\cdot\:]$ does not depend on $\alpha$.
For this 1D EM field, here the decay rate of the TLS is $\gamma=\wp^{2}\,\Omega_{\mathsf{eg}}\big/\hbar\epsilon_{0}\mathsf{S}c$
(see Appendix \ref{sec:quantum}). 

\subsection{Bloch equation solution: zero driving phase \label{subsec:Bloch-0-phase}}

Once the solution for the above master equation (\ref{eq:TLS-ME})
is obtained (either analytically or numerically), the \emph{P}-function
average of $\tilde{\rho}^{(\alpha)}_{\text{\textsc{s}}}(t)$ would
give the full system dynamics {[}Eq.\,(\ref{eq:unitary}){]} driven
by nonclassical light pulses. 

Generally, when we need to consider both the TLS dissipation and the
time dependent pulse driving, it is not easy to solve the above master
equation (\ref{eq:TLS-ME}) analytically \citep{allen_optical_1987,zlatanov_exact_2015,grira_atomic_2021,grira_exact_2023}.
But here we find that, if the light pulse has the above the exponential
shape (\ref{eq:E(xt)}), in the resonance case ($\Omega_{\text{eg}}=\omega_{\text{d}}$),
the TLS dynamics can be analytically solved. 

We first consider the situation that $\alpha$ is a positive real
number, thus $\mathcal{E}_{\alpha}=\alpha\,\eta_{0}/\hbar>0$. In
this case, the above driving interaction (\ref{eq:V_a-TLS}) further
becomes $\tilde{V}_{\alpha}(t)=-\frac{1}{2}\hbar\mathcal{E}_{\alpha}\,\hat{\sigma}^{x}\,e^{-\frac{1}{2}\kappa t}$.
Denoting $u=\langle\hat{\sigma}^{x}\rangle$, $v=\langle\hat{\sigma}^{y}\rangle$,
and $w=\langle\hat{\sigma}^{z}\rangle$, the master equation (\ref{eq:TLS-ME})
gives the following Bloch equations (here $\langle\hat{\sigma}^{x,y,z}(t)\rangle$
are expectations in the $\alpha$-branch, but we temporally omit their
superscripts ``$\alpha$'' for simplicity): \begin{subequations}
\begin{align}
\frac{\text{d}u}{\text{d}t} & =-\frac{1}{2}\gamma\,u,\label{eq:bloch-u}\\
\frac{\text{d}v}{\text{d}t} & =-\frac{1}{2}\gamma\,v+\mathcal{E}_{\alpha}\,e^{-\frac{1}{2}\kappa t}\,w(t),\label{eq:bloch-v}\\
\frac{\text{d}w}{\text{d}t} & =-\gamma\,(w+1)-\mathcal{E}_{\alpha}\,e^{-\frac{1}{2}\kappa t}\,v(t).\label{eq:bloch-w}
\end{align}
\end{subequations} In such a scattering problem, we consider initially
the TLS is in the ground state $|\mathsf{g}\rangle$, which sets the
initial conditions as $u(0)=v(0)=0$, and $w(0)=-1$.

The first equation (\ref{eq:bloch-u}) simply gives $u(t)=u(0)\,e^{-\frac{1}{2}\gamma t}=0$.
Now we need to solve the coupled equations for $v(t)$ and $w(t)$.
Here we make a variable transformation $\zeta_{t}:=e^{-\frac{1}{2}\kappa t}\in(0,1]$
\citep{allen_optical_1987,zlatanov_exact_2015}, then Eqs.\,(\ref{eq:bloch-v},
\ref{eq:bloch-w}) lead to the following second order differential
equation (see Appendix \ref{sec:classical}),
\begin{equation}
\zeta^{2}\frac{\text{d}^{2}\bar{w}}{\text{d}\zeta^{2}}-3\tilde{\gamma}\zeta\frac{\text{d}\bar{w}}{\text{d}\zeta}+\big[4\tilde{\mathcal{E}}^{2}\zeta^{2}+2\tilde{\gamma}(\tilde{\gamma}+1)\big]\bar{w}=4\tilde{\mathcal{E}}^{2}\zeta^{2},\label{eq:wbar(zeta)}
\end{equation}
Here $\bar{w}(\zeta):=w(\zeta)+1=2\bar{\text{\textsc{n}}}_{\mathsf{e}}$
($\bar{\text{\textsc{n}}}_{\mathsf{e}}$ is the excited population),
and $\tilde{\gamma}:=\gamma/\kappa$ and $\tilde{\mathcal{E}}:=\mathcal{E}_{\alpha}/\kappa$
are unitless.

Once the solution of $\bar{w}(t)=w(t)+1$ is obtained, the evolution
of $v(t)$ can be immediately solved exactly from Eq.\,(\ref{eq:bloch-v}).
By adopting the Fr\"obenius method, setting $\bar{w}(\zeta):=\zeta^{p}\,y(\zeta)$,
Eq.\,(\ref{eq:wbar(zeta)}) gives an inhomogeneous Bessel equation
of $y(\zeta)$, and finally the exact analytical solution of $\bar{w}(\zeta)$
is obtained as (see Appendix \ref{sec:classical}) \begin{widetext}
\begin{equation}
\begin{split}\bar{w}^{(\alpha)}\big(\zeta\equiv e^{-\frac{1}{2}\kappa t}\big)=1+w^{(\alpha)}(\zeta) & =2\pi\,\tilde{\mathcal{E}}^{2}\,\zeta^{p}\int^{\zeta}_{1}\Big[Y_{\nu}(2\tilde{\mathcal{E}}\zeta)J_{\nu}(2\tilde{\mathcal{E}}\eta)-J_{\nu}(2\tilde{\mathcal{E}}\zeta)Y_{\nu}(2\tilde{\mathcal{E}}\eta)\Big]\eta^{1-p}\,d\eta,\\
p & =\frac{1}{2}(1+3\tilde{\gamma}),\qquad\nu=\frac{1}{2}|\tilde{\gamma}-1|.
\end{split}
\label{eq:wb_final}
\end{equation}
\end{widetext} Here, we put back the superscript ``$\alpha$''
for the $\alpha$-branch. $J_{\nu}(x)$ and $Y_{\nu}(x)$ are Bessel
functions. Initially the atom stays in the ground state ($\langle\hat{\sigma}^{z}\rangle_{t=0}=-1$),
then after the pulse excitation, it would return the ground state
after the spontaneous emission ($\langle\hat{\sigma}^{z}\rangle_{t\rightarrow\infty}=-1$).
That gives $\bar{w}\big|_{\zeta=0}=\bar{w}\big|_{\zeta=1}=0$ as the
boundary condition, and the above integration form and the Bessel
function properties around $\zeta=0$ guarantee this boundary condition
is well satisfied. 

As a simple verification, in the limit of no dissipation ($\tilde{\gamma}=0$
and $p=\nu=1/2$), we have $J_{1/2}(x)=\sqrt{2/(\pi x)}\,\sin x$
and $Y_{1/2}(x)=-\sqrt{2/(\pi x)}\,\cos x$, and the above result
(\ref{eq:wb_final}) gives
\begin{equation}
\bar{\text{\textsc{n}}}^{(\alpha)}_{\mathsf{e}}(t)=\frac{1}{2}\bar{w}^{(\alpha)}(\zeta_{t})=\sin^{2}\big[\frac{\mathcal{E}_{\alpha}}{\kappa}(1-e^{-\frac{1}{2}\kappa t})\big].\label{eq:Pe(t)}
\end{equation}
 This result also can be obtained and verified by directly solving
the Bloch equations (\ref{eq:bloch-u}-\ref{eq:bloch-w}) when $\gamma=0$.

\subsection{General nonzero driving phases }

Now we study a more general case that the driving strength $\mathcal{E}_{\alpha}\equiv\alpha\,\eta_{0}/\hbar$
is a general complex number ($\alpha\equiv|\alpha|\,e^{i\phi_{\alpha}}$).
In this case, the above driving interaction (\ref{eq:V_a-TLS}) now
becomes ($\Omega_{\mathsf{eg}}=\omega_{\text{d}}$)
\begin{equation}
\tilde{V}_{\alpha}=-\frac{1}{2}\hbar|\mathcal{E}_{\alpha}|\,(e^{i\phi_{\alpha}}\,\hat{\sigma}^{+}+e^{-i\phi_{\alpha}}\,\hat{\sigma}^{-})\,e^{-\frac{1}{2}\kappa t}.
\end{equation}

To study such cases, we define $\hat{\sigma}^{+}_{\phi}:=e^{i\phi_{\alpha}}|\mathsf{e}\rangle\langle\mathsf{g}|\equiv|\mathsf{e}_{\phi}\rangle\langle\mathsf{g}|$,
which absorbs the driving phase in $|\mathsf{e}_{\phi}\rangle\equiv e^{i\phi_{\alpha}}|\mathsf{e}\rangle$.
Then the above interaction becomes $\tilde{V}_{\alpha}=-\frac{1}{2}\hbar|\mathcal{E}_{\alpha}|\,(\hat{\sigma}^{+}_{\phi}+\hat{\sigma}^{-}_{\phi})\,e^{-\frac{1}{2}\kappa t}$,
which returns the same form as the above zero phase case (Sec.\,\ref{subsec:Bloch-0-phase}),
and all the above results directly apply. 

Now we consider how these expectations with/without the phase $\phi_{\alpha}$
can be transformed between each other. For the expectations $\langle\hat{\sigma}^{x,y,z}\rangle$
and $\langle\hat{\sigma}^{x,y,z}_{\phi}\rangle$ defined by $\{|\mathsf{e}\rangle,\,|\mathsf{g}\rangle\}$
and $\{|\mathsf{e}_{\phi}\rangle,\,|\mathsf{g}\rangle\}$, they satisfy
the relation 
\begin{align}
\langle\hat{\sigma}^{+}\rangle & =\frac{1}{2}(\langle\hat{\sigma}^{x}\rangle+i\langle\hat{\sigma}^{y}\rangle)=e^{-i\phi_{\alpha}}\langle\hat{\sigma}^{+}_{\phi}\rangle\nonumber \\
 & =\frac{1}{2}(\cos\phi_{\alpha}-i\sin\phi_{\alpha})(\langle\hat{\sigma}^{x}_{\phi}\rangle+i\langle\hat{\sigma}^{y}_{\phi}\rangle).
\end{align}
Collecting the real and imaginary parts gives 
\begin{equation}
\begin{split}\langle\hat{\sigma}^{x}\rangle & =\langle\hat{\sigma}^{x}_{\phi}\rangle\cos\phi_{\alpha}+\langle\hat{\sigma}^{y}_{\phi}\rangle\sin\phi_{\alpha}\\
\langle\hat{\sigma}^{y}\rangle & =-\langle\hat{\sigma}^{x}_{\phi}\rangle\sin\phi_{\alpha}+\langle\hat{\sigma}^{y}_{\phi}\rangle\cos\phi_{\alpha},\quad\langle\hat{\sigma}^{z}\rangle=\langle\hat{\sigma}^{z}_{\phi}\rangle.
\end{split}
\label{eq:phase-transform}
\end{equation}

Therefore, to study the situation with a nonzero driving phase $\phi_{\alpha}\neq0$,
we could first consider expectations $\langle\hat{\sigma}^{x,y,z}_{\phi}(t)\rangle$.
The driving phase is formally eliminated, and the dynamical equations
just return the above Bloch equations (\ref{eq:bloch-u}-\ref{eq:bloch-w})
for the zero-phase situation, which have been well solved. Then the
expectations $\langle\hat{\sigma}^{x,y,z}(t)\rangle$ can be obtained
by the transformation (\ref{eq:phase-transform}). 

It is worth noting that, for $\hat{\sigma}^{z}=|\mathsf{e}\rangle\langle\mathsf{e}|-|\mathsf{g}\rangle\langle\mathsf{g}|$,
we always have $\hat{\sigma}^{z}=\hat{\sigma}^{z}_{\phi}$ and $\langle\hat{\sigma}^{z}(t)\rangle=\langle\hat{\sigma}^{z}_{\phi}(t)\rangle$
for arbitrary $\phi_{\alpha}\neq0$. That means $\langle\hat{\sigma}^{z}(t)\rangle$
only depends on the light intensity ($\sim|\mathcal{E}_{\alpha}|^{2}$)
but independent on the driving phase. Therefore, for the quantity
$\bar{w}(t)=1+\langle\hat{\sigma}^{z}(t)\rangle$, we can directly
substitute $\mathcal{E}_{\alpha}\equiv\alpha\,\eta_{0}/\hbar$ as
$|\mathcal{E}_{\alpha}|$ in the above Eq.\,(\ref{eq:wbar(zeta)}),
and that well gives the results for the nonzero driving phase situations. 

Therefore, for general driving phases, it turns out the above integration
form (\ref{eq:wb_final}) can be expanded as a summation series form
$\bar{w}^{(\alpha)}(\zeta)=\sum^{\infty}_{K=1}\,\mathscr{C}_{K}(\zeta)\,|\tilde{\mathcal{E}}|^{2K}$
with respect to the driving strength $\tilde{\mathcal{E}}\equiv\alpha\,\eta_{0}/\hbar\kappa$,
that is (see Appendix \ref{sec:classical}), 
\begin{align}
\bar{w}^{(\alpha)}(\zeta) & =\sum^{\infty}_{K=1}\,\mathscr{C}_{K}(\zeta)\,\big(\frac{\eta_{0}}{\hbar\kappa}\big)^{2K}\cdot|\alpha|^{2K},\nonumber \\
\mathscr{C}_{K}(\zeta) & =\frac{2}{\nu}\sum^{K-1}_{q=0}\frac{(-1)^{K-1}\Gamma(1+\nu)\Gamma(1-\nu)\,\zeta^{2K}}{q!(K-1-q)!\Gamma(1+q+\nu)\Gamma(K-q-\nu)}\nonumber \\
 & \times\Big(\frac{1-\zeta^{p+\nu+2q-2K}}{2K-2q-p-\nu}-\frac{1-\zeta^{p-\nu-2q-2}}{2+2q-p+\nu}\Big).\label{eq:wb-expand}
\end{align}

For the 1D EM field here with the resonant condition $\Omega_{\mathsf{eg}}=\omega_{\text{d}}$,
the TLS decay rate $\gamma$, the pulse linewidth $\kappa$, and the
single-photon coupling strength $\eta_{0}$ have a critical relation
$\eta_{0}=\hbar\sqrt{2\gamma\kappa}$ (see their definitions and Appendix
\ref{sec:quantum}), which is quite useful for the discussions below.

If the pulse shape is not the above exponential form (\ref{eq:E(xt)}),
or the the frequencies are not resonant $\Omega_{\mathsf{eg}}\neq\omega_{\text{d}}$,
we may not have the analytical results for the observable expectations
$\langle\hat{O}^{(\alpha)}_{\text{\textsc{s}}}(t)\rangle$ in each
quasi-classical $\alpha$-branch, while in principle they can be obtained
numerically. Now the problem is turned into how to deal with the specific
\emph{P}-function average for the nonclassical input light. 

\section{Applications: interacting with a nonclassical multi-photon pulse
\label{sec:nonclassical}}

With the above preparations, now we study the TLS dynamics when it
is interacting with different kinds of light states, especially the
nonclassical states, such as the Fock states and squeezed vacuum state.

\subsection{Benchmark: interacting with the single-photon pulse \label{subsec:Benchmark-1p}}

We first study the situation that the input light is a single-photon
pulse, which is a typical nonclassical light state. In this case,
the light state is written as $|\Psi_{1\text{p}}\rangle=\sum_{k}\,\beta_{k}\,\hat{a}^{\dagger}_{k}|\varnothing\rangle$
\citep{loudon_quantum_2000,liao_spectrum_2012,liao_single-photon_2015,Liao2020},
and here we consider $\{\beta_{k}\}$ takes the above Lorentzian form
(\ref{eq:beta-k}). This state also can be represented by the above
pulse-shaped \emph{P}-function (\ref{eq:P-pulse}) with (see Appendix
\ref{sec:Apx-pulse})
\begin{equation}
P_{1\text{p}}(\alpha,\alpha^{*})=e^{|\alpha|^{2}}\frac{\partial^{2}}{\partial\alpha\partial\alpha^{*}}\delta^{(2)}(\alpha).
\end{equation}

Based on the above algorithm in Sec.\,\ref{subsec:nonclassical},
when the input light is such a single-photon pulse, the TLS behavior
$\bar{w}_{1\text{p}}(t)$ now can be obtained by taking the \emph{P}-function
average (\ref{eq:O-average}) from all $\alpha$-branch evolutions
$\bar{w}^{(\alpha)}(\zeta)$ {[}Eqs.\,(\ref{eq:wb_final}, \ref{eq:wb-expand}){]},
and that gives 
\begin{align}
\bar{w}_{1\text{p}} & (\zeta_{t})=\int\text{d}^{2}\alpha\,P_{1\text{p}}(\alpha,\alpha^{*})\,\bar{w}^{(\alpha)}(\zeta_{t})\nonumber \\
 & =\partial^{N}_{\alpha}\partial^{N}_{\alpha^{*}}\big[e^{|\alpha|^{2}}\bar{w}^{(\alpha)}(\zeta)\big]\big|_{\alpha\rightarrow0}=\mathscr{C}_{1}(\zeta)\,\big(\frac{\eta_{0}}{\hbar\kappa}\big)^{2}.\label{eq:w1p}
\end{align}
 The last equality utilizes the expansion form (\ref{eq:wb-expand}),
where $\mathscr{C}_{K=1}(\zeta)$ is the expansion coefficient (see
Appendix \ref{sec:classical}). As a result, when the TLS is interacting
with a single-photon pulse, the full evolution of the excited population
is ($\eta_{0}=\hbar\sqrt{2\gamma\kappa}$)
\begin{equation}
\bar{\text{\textsc{n}}}^{(1\text{p})}_{\mathsf{e}}(t)=\frac{1}{2}\bar{w}_{1\text{p}}(\zeta_{t})=\frac{\eta^{2}_{0}}{\hbar^{2}}\cdot\frac{(e^{-\frac{1}{2}\gamma t}-e^{-\frac{1}{2}\kappa t})^{2}}{(\gamma-\kappa)^{2}}.\label{eq:Ne(t)-1p}
\end{equation}

Particularly, in the limit $\gamma=\kappa$, this excitation probability
becomes $\bar{\text{\textsc{n}}}_{\mathsf{e}}(t)=\frac{1}{2}\gamma^{2}t^{2}\,e^{-\gamma t}$,
and its maximum is $\bar{\text{\textsc{n}}}^{\text{max}}_{\mathsf{e}}=2/e^{2}\approx0.27$
at $\gamma t=2$ {[}see the solid blue line in Fig.\,\ref{fig-Fock}(a){]}. 

Indeed, this problem with the single-photon pulse input also can be
analytically solved by the fully quantum formalism. In this fully
quantum treatment, after RWA, $\{|\mathsf{e}\rangle\otimes|\varnothing\rangle,\,|\mathsf{g}\rangle\otimes\hat{a}^{\dag}_{k}|\varnothing\rangle\}$
forms a closed single-excitation subspace for the whole system-field
evolution, and the TLS dynamics can be analytically solved from the
Schr\"odinger equation. It turns out this approach exactly gives
the same result as the above Eq.\,(\ref{eq:Ne(t)-1p}) obtained from
the \emph{P}-function average approach (see details in Appendix \ref{sec:quantum}). 

This specific example with the single-photon pulse input well confirms
the validity of the above general framework about interacting with
a generic nonclassical light pulse as discussed in Sec.\,\ref{sec:General}.

\subsection{Interacting with the \emph{N}-photon Fock pulse }

Now we further consider a more complicated situation that the input
light is the pulse-shaped Fock state $|N,\{\beta_{k}\}\rangle$. Again,
we write down the light state by the above pulse-shaped \emph{P}-function
(\ref{eq:P-pulse}), where $\{\beta_{k}\}$ takes the Lorentzian form
(\ref{eq:beta-k}), and the \emph{P}-function is (see Appendix \ref{sec:Apx-pulse})
\citep{vogel_quantum_2006,walls_quantum_2008,agarwal_quantum_2012}
\begin{equation}
P_{|N\rangle}(\alpha,\alpha^{*})=\frac{e^{|\alpha|^{2}}}{N!}\partial^{N}_{\alpha}\partial^{N}_{\alpha^{*}}\delta^{(2)}(\alpha).
\end{equation}
 Clearly, $P_{|N\rangle}(\alpha,\alpha^{*})$ is a highly singular,
generalized function. 

Similarly as the above discussions, when interacting with such a nonclassical
Fock pulse, the evolution of the excited population becomes 
\begin{align}
\bar{\text{\textsc{n}}}^{|N\rangle}_{\mathsf{e}}(t) & =\int\text{d}^{2}\alpha\,P_{|N\rangle}(\alpha,\alpha^{*})\cdot\frac{1}{2}\bar{w}^{(\alpha)}(\zeta_{t})\nonumber \\
 & =\frac{1}{2}\cdot\frac{1}{N!}\partial^{N}_{\alpha}\partial^{N}_{\alpha^{*}}\big[\,e^{|\alpha|^{2}}\,\bar{w}^{(\alpha)}(\zeta)\,\big]\big|_{\alpha\rightarrow0}\nonumber \\
 & =\frac{1}{2}\cdot\sum^{N}_{K=1}\mathscr{C}_{K}(\zeta)\,\big(\frac{\eta_{0}}{\hbar\kappa}\big)^{2K}\cdot\frac{N!}{(N-K)!}.\label{eq:Fock-avg}
\end{align}
Here the result $\bar{w}^{(\alpha)}(\zeta_{t})$ under the coherent
pulse {[}Eqs.\,(\ref{eq:wb_final}, \ref{eq:wb-expand}){]} serves
as a generating function, from which the dynamics driven by the Fock
pulse can be efficiently evaluated.

\begin{figure}
\includegraphics[width=1\columnwidth]{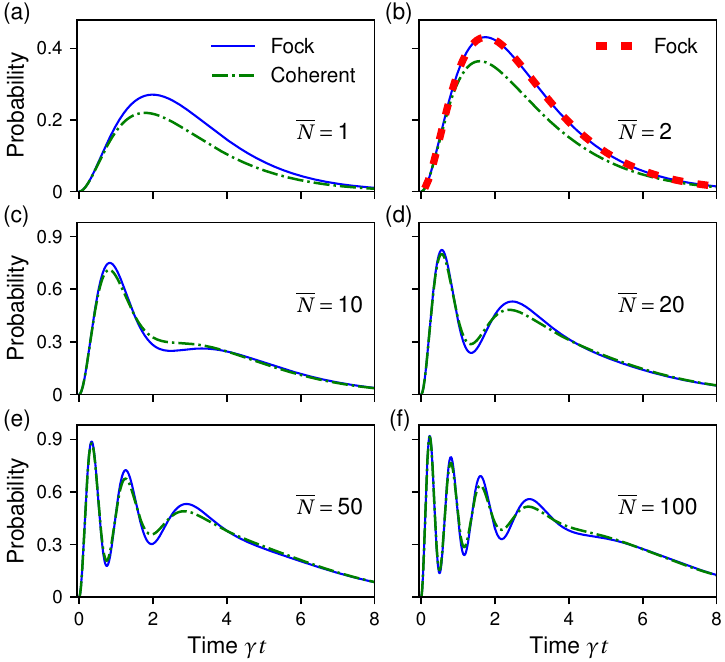}

\caption{The excitation probability evolution $\bar{\textsc{n}}_{\mathsf{e}}(t)$
from different input pulse states with total photon number $\overline{N}$.
The dot-dashed green lines show the situation when the input pulse
is the coherent state $|\alpha,\{\beta_{k}\}\rangle$ with $|\alpha|^{2}=\overline{N}$
as the total average photon number {[}from Eq.\,(\ref{eq:wb_final}){]}.
The solid blue lines show the situation when the input pulse is the
Fock state $|\overline{N},\{\beta_{k}\}\rangle$ and are calculated
by Eq.\,(\ref{eq:Fock-avg}). The dashed red line in (b) also shows
the situation of Fock input for comparison, which is calculated by
the generalized master equation method \citep{Gheri1998,Shi2015,You2018,Liao2020,yang_quantum_2024}.
The pulse shape $\{\beta_{k}\}$ takes the Lorentzian distribution
(\ref{eq:beta-k}), and here we consider $\gamma=\kappa$.}
 \label{fig-Fock}
\end{figure}

Comparing with the series form (\ref{eq:wb-expand}) for the quasi-classical
driving situation, the result (\ref{eq:Fock-avg}) driven by the Fock
pulse has a quite similar form, where the power terms of the photon
numbers $|\alpha|^{2K}$ are substituted as $N!/(N-K)!=\langle N|\hat{a}^{\dag K}\hat{a}^{K}|N\rangle$.
Indeed, this can be simply understood from the following property
of \emph{P}-function, 
\begin{equation}
\int\text{d}^{2}\alpha\,P(\alpha)\,|\alpha|^{2K}=\langle:\hat{a}^{\dag K}\hat{a}^{K}:\rangle=g^{(K)}\,\langle\hat{a}^{\dag}\hat{a}\rangle^{K},
\end{equation}
where $\langle:f(\hat{a}^{\dag},\hat{a}):\rangle$ indicates the normal
order expectation, and $g^{(K)}:=\langle\hat{a}^{\dag K}\hat{a}^{K}\rangle/\langle\hat{a}^{\dag}\hat{a}\rangle^{K}$
is the $K$-order optical coherence function \citep{cao_quantum_2022,yao_enhancing_2020}. 

Therefore, comparing with the series form (\ref{eq:wb-expand}) for
the observable expectation in the quasi-classical driving situation,
when the system is interacting with a nonclassical light, we can simply
replace the power terms $|\alpha|^{2K}$ of the light intensity by
$g^{(K)}\langle\hat{a}^{\dag}\hat{a}\rangle^{K}$, and that would
formally give the result for different input light states \citep{cao_quantum_2022,yao_enhancing_2020}.
In more general cases, we may not have the analytical results for
$w^{(\alpha)}(\zeta_{t})$, then we need to solve the quasi-classical
driving master equation numerically, as well as the derivatives in
Eq.\,(\ref{eq:Fock-avg}). 

In Fig.\,\ref{fig-Fock}, we show the evolution of the excitation
probability $\bar{\text{\textsc{n}}}_{\mathsf{e}}(t)$ when the input
light is the Fock pulse with different photon umbers $N=1,2,10,20,50,100$
{[}solid blue lines, obtained by Eq.\,(\ref{eq:Fock-avg}){]}, in
comparison with the situations of coherent input with the same average
photon numbers {[}dot-dashed green lines, obtained by Eq.\,(\ref{eq:wb_final}){]}.
When the mean photon number $\overline{N}$ is large, the difference
between the situations of Fock input and coherent input looks relatively
smaller. This can be understood by the above expansion (\ref{eq:Fock-avg})
corrected by $g^{(K)}$. For the Fock states with a large $N$, the
$K$-order coherences $g^{(K)}_{|N\rangle}=N!\big/[(N-K)!\,N^{K}]$
are quite close to $1$ for the heading orders $K\ll N$, thus they
would give a result quite similar as the situation of the coherent
input ($g^{(K)}=1$ for all $K$). 

In Fig.\,\ref{fig-Fock}(b), we show the evolution of $\bar{\text{\textsc{n}}}_{\mathsf{e}}(t)$
when the input light is the two-photon pulse $|N=2,\{\beta_{k}\}\rangle$,
and we adopt the generalized master equation method (Refs.\,\citep{Gheri1998,Shi2015,You2018,Liao2020,yang_quantum_2024})
to study this problem for comparison (the dashed red line). It turns
out our result by \emph{P}-function average (solid blue line) shows
perfect agreement with the generalized master equation approach, which
well confirms the validity of our approach. 

\subsection{The light pulses with more general photon statistics}

Here we study some more general situations when the input light pulse
has certain specific photon statistics. We first consider the situation
that the input light is a ``thermal'' pulse. The light state is
represented by the pulse-shaped \emph{P}-function (\ref{eq:P-pulse}),
where $\{\beta_{k}\}$ gives the pulse shape, and 
\begin{equation}
P_{\text{th}}(\alpha,\alpha^{*})=\frac{1}{\pi\bar{\mathsf{n}}_{\text{th}}}\exp\big(-\frac{|\alpha|^{2}}{\bar{\mathsf{n}}_{\text{th}}}\big).
\end{equation}
Here $\bar{\mathsf{n}}_{\text{th}}$ is the average photon number
in this light pulse in total. Equivalently, it can be verified that
such a state also can be represented by the density state form 
\begin{gather}
\hat{\boldsymbol{\rho}}^{\text{(th)}}_{\text{\textsc{e}}}=\sum^{\infty}_{N=0}p^{\text{(th)}}_{N}\,\big|N,\{\beta_{k}\}\big\rangle\big\langle N,\{\beta_{k}\}\big|,\label{eq:th-pulse}\\
p^{\text{(th)}}_{N}=\frac{1}{1+\bar{\mathsf{n}}_{\text{th}}}\big(\frac{\bar{\mathsf{n}}_{\text{th}}}{1+\bar{\mathsf{n}}_{\text{th}}}\big)^{N}.\nonumber 
\end{gather}
Here, each component $\big|N,\{\beta_{k}\}\big\rangle$ is the pulse-shaped
Fock state, and the probability $p^{\text{(th)}}_{N}$ follows the
exponential distribution. In this sense, we call it a ``thermal''
pulse.

When the TLS is interacting with such a thermal pulse, the TLS evolution
can be calculated by taking the \emph{P}-function average on the integral
form (\ref{eq:wb_final}) of the solution $\bar{w}^{(\alpha)}(\zeta_{t})$.
Or we can take the \emph{P}-function average from the series form
(\ref{eq:wb-expand}), and that formally gives, 
\begin{align}
\bar{\text{\textsc{n}}}^{\text{(th)}}_{\mathsf{e}}(t) & =\int\text{d}^{2}\alpha\,P_{\text{th}}(\alpha,\alpha^{*})\cdot\frac{1}{2}\bar{w}^{(\alpha)}(\zeta_{t})\nonumber \\
 & =\frac{1}{2}\cdot\sum^{\infty}_{K=1}\mathscr{C}_{K}(\zeta)\,\big(\frac{\eta_{0}}{\hbar\kappa}\big)^{2K}\int\text{d}^{2}\alpha\,\frac{1}{\pi\bar{\mathsf{n}}_{\text{th}}}e^{-\frac{|\alpha|^{2}}{\bar{\mathsf{n}}_{\text{th}}}}|\alpha|^{2K}\nonumber \\
 & =\frac{1}{2}\cdot\sum^{\infty}_{K=1}\mathscr{C}_{K}(\zeta)\,\big(\frac{\eta_{0}}{\hbar\kappa}\big)^{2K}\cdot K!\,\bar{\mathsf{n}}^{K}_{\text{th}}.
\end{align}

As mentioned above, here the power terms of the average photon numbers
$|\alpha|^{2K}$ are corrected as $K!\,\bar{\mathsf{n}}^{K}_{\text{th}}$,
where $g^{(K)}_{\text{th}}=K!$ is the $K$-order optical coherence
for the thermal state. 

\begin{figure}
\includegraphics[width=1\columnwidth]{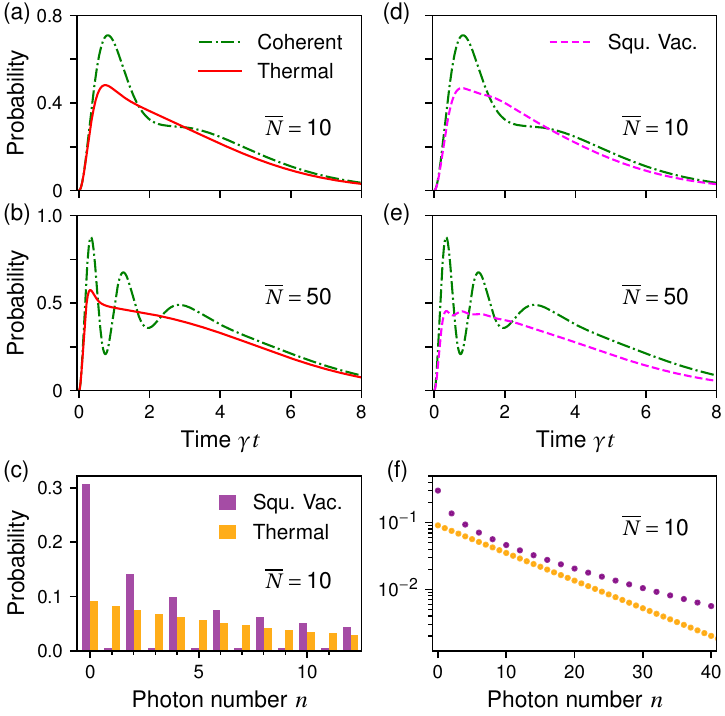}

\caption{The excitation probability evolution $\bar{\textsc{n}}_{\mathsf{e}}(t)$
driven by the thermal or squeezed pulses ($\gamma=\kappa$). The dot-dashed
green lines in (a, b, d, e): the input light is the coherent pulse
for comparison. The solid red lines in (a, b): the input light is
the thermal pulse (\ref{eq:th-pulse}). The dashed purple lines in
(d, e): the input pulse follows the photon statistics of the squeezed
vacuum state (\ref{eq:Pn-sqv}). (c, f) Demonstration for the photon
statistics $\{p_{n}\}$ for the thermal state and the squeezed vacuum
state (mean photon number $\overline{N}=10$, corresponding to a squeezing
strength $r\simeq1.87$). The conditions are the same with Fig.\,\ref{fig-Fock}.}
 \label{fig-th}
\end{figure}

As another method, here we can also utilize the above result (\ref{eq:Fock-avg})
for the Fock pulse input, and that gives 
\begin{align}
\bar{\text{\textsc{n}}}^{\text{(th)}}_{\mathsf{e}}(t) & =\sum^{\infty}_{N=0}\,p^{\text{(th)}}_{N}\cdot\bar{\text{\textsc{n}}}^{|N\rangle}_{\mathsf{e}}(t),\label{eq:Ne(t)-sum_N}
\end{align}
where $\bar{\text{\textsc{n}}}^{|N\rangle}_{\mathsf{e}}(t)$ is the
population evolution when the input light is the Fock pulse. 

More generally, if the input light pulse can be represented as $\hat{\boldsymbol{\rho}}_{\text{\textsc{e}}}=\sum\,p_{N}\,\big|N,\{\beta_{k}\}\big\rangle\big\langle N,\{\beta_{k}\}\big|$,
the TLS evolution also can be obtained similarly as the above Eq.\,(\ref{eq:Ne(t)-sum_N}).
Here the $\{p_{N}\}$ is the photon statistics of the input light,
which could be an arbitrary type, such as the nonclassical statistics
\citep{cao_quantum_2022,yao_enhancing_2020}. 

For example, for the squeezed vacuum (SqV) state, the photon statistics
$\{p_{N}\}$ is \citep{agarwal_quantum_2012,walls_quantum_2008}
\begin{equation}
p^{\text{(sqv)}}_{2n}=\frac{(2n-1)!!}{(2n)!!}\frac{(\tanh r)^{2n}}{\cosh r},\hspace*{1em}n=0,1,2,\dots\label{eq:Pn-sqv}
\end{equation}
and all the odd terms are zero $p^{\text{(sqv)}}_{2n+1}=0$ {[}Fig.\,\ref{fig-th}(e,
f){]}. This is a highly nonclassical light state (mean photon number
$\langle\hat{a}^{\dag}\hat{a}\rangle=\sinh^{2}r$). 

In Fig.\,\ref{fig-th}, we show the evolution of $\bar{\text{\textsc{n}}}_{\mathsf{e}}(t)$
when the input light follows the thermal statistics {[}Eq.\,(\ref{eq:th-pulse}),
see solid red lines{]}, or the SqV statistics {[}Eq.\,(\ref{eq:Pn-sqv}),
see dashed purple lines{]}, in comparison with the results of coherent
input (dot-dashed green lines) with the same average photon number
$\overline{N}$. For the situations of Fock input and coherent input
(Fig.\,\ref{fig-Fock}), when the photon number in the input pulse
is large, the excitation probability $\bar{\text{\textsc{n}}}_{\mathsf{e}}(t)$
exhibits significant oscillations, but such oscillations do not appear
in the situations of thermal pulse and SqV input (Fig.\,\ref{fig-th}). 

Such a behavior can be understood by the following picture. First,
clearly these oscillations can be understood as similarly to the Rabi
oscillation in the quasi-classical driving picture, and the frequency
of these Rabi oscillations are determined by the input light intensity
or the photon number $N$. For the thermal and SqV input situations,
the full evolution of $\bar{\text{\textsc{n}}}_{\mathsf{e}}(t)$ can
be regarded as the probabilistic average of all different branches
$\bar{\text{\textsc{n}}}^{|N\rangle}_{\mathsf{e}}(t)$ from the Fock
inputs (\ref{eq:Ne(t)-sum_N}). The oscillation frequencies in different
Fock branches are determined by the input photon number $N$; meanwhile,
the photon statistics $p_{N}$ for the thermal and SqV statistics
are relatively flat and wide. Thus, in the final result of $\bar{\text{\textsc{n}}}_{\mathsf{e}}(t)$,
the average of all these asynchronous oscillations in different Fock
branches cancels such an oscillating behavior. 

\section{Conclusion \label{sec:conclusion}}

In sum, we have developed a rigorous and versatile theoretical framework
to describe the quantum dynamics of a multi-level system interacting
with an arbitrary nonclassical light pulse. By introducing the pulse-shaped
\emph{P}-representation, the full quantum evolution can be decomposed
and regarded as a ``probabilistic'' mixture of many quasi-classical
``branches''. In each evolution branch, the system dynamics can
be well described by the quasi-classical master equation widely adopted
in literature. Finally, the evolution of the system state and observable
expectations can be obtained as the \emph{P}-function average from
these quasi-classical evolution branches.

To be more specific, here we study the dynamics of a TLS interacting
with a nonclassical light pulse. For the widely used exponential light
pulse, in each quasi-classical branch, we derived exact analytical
solutions for the TLS dynamics in terms of Bessel functions. Then
this branch-averaging method directly gives the analytical results
driven by nonclassical Fock pulses with an arbitrary photon number.
For the single-photon and two-photon cases, our method shows perfect
agreement with the existing results in literature (analytically or
numerically), which well confirms the validity of this approach. This
allows for the efficient simulation involving a nonclassical pulse
state with a large photon number (e.g., $N\sim100$ Fock states),
and more complex states, such as the thermal states, and SqV states. 

This approach effectively separates the quantum statistical nature
of the light field (captured by the \emph{P}-function) from its temporal
pulse-shape effects, and well utilizes the quasi-classical master
equation widely adopted in literature. In this sense, rather than
relying on direct expansions, the present framework can be understood
as a generating-function approach, in which the full quantum dynamics
is obtained by taking proper derivative operations on the quasi-classical
branches. This work provides a unified tool for exploring how the
quantum statistics of light can be used to control and manipulate
matter at the quantum level. Given its generality, this framework
can be readily extended to more complex systems.

\vspace{1em}

\emph{Acknowledgments }- SWL is grateful to L.-P. Yang for the illuminating
discussions on the quantum description of light pulses, and for the
insights into solving the Bloch equations. This study is supported
by Innovation Program for Quantum Science and Technology (Grant No.\,2023ZD0300700),
NSF of China (Grants No.\,12475030 and No.\,12475033), Guangdong
Provincial Quantum Science Strategic Initiative (Grant No.\,GDZX2505001),
Guangdong Basic and Applied Basic Research Foundation (Grant No.\,2026A1515011705).

\appendix
\begin{widetext}

\section{The\emph{ P}-function for a generic nonclassical light pulse \label{sec:Apx-pulse} }

Here we study how to write down the quantum state of a nonclassical
light pulse in free space by the \emph{P}-representation \citep{schleich_quantum_2001,vogel_quantum_2006,walls_quantum_2008,agarwal_quantum_2012}.
First, we introduce the collective pulse photon operator $\hat{A}:=\sum_{k}\,\beta_{k}\,\hat{a}_{k}$
\citep{loudon_quantum_2000,Liao2020}. It is simple to verify $[\hat{A},\hat{A}^{\dag}]=1$,
which means they satisfy the same relations as the boson operator
$\hat{a},\,\hat{a}^{\dag}$ for a single optical mode. For example,
the quantum state for a single photon pulse in free space can be written
as \citep{loudon_quantum_2000,Liao2020} 
\begin{equation}
\big|\Psi_{1\mathrm{p}}\big\rangle=\hat{A}^{\dag}|\varnothing\rangle=\sum_{k}\beta_{k}\,\hat{a}^{\dag}_{k}|\varnothing\rangle=\sum_{k}\beta_{k}\,|1_{k}\rangle,\qquad|1_{k}\rangle:=\hat{a}^{\dag}_{k}|\varnothing\rangle,\label{eq:1-photon}
\end{equation}
 where $|\beta_{k}|^{2}$ is the probability to find a single photon
in mode-$k$. More generally, we can further define the Fock pulse
state with $N$ photons, that is \citep{loudon_quantum_2000,Liao2020},
\begin{equation}
|N,\{\beta_{k}\}\rangle=\frac{(\hat{A}^{\dag})^{N}}{\sqrt{N!}}\,|\varnothing\rangle,\qquad\langle N,\{\beta_{k}\}|\,\hat{a}^{\dag}_{q}\hat{a}_{q}\,|N,\{\beta_{k}\}\rangle=\frac{1}{N!}\langle\varnothing|\hat{A}^{N}\,\hat{a}^{\dag}_{q}\hat{a}_{q}\,\hat{A}^{\dag N}|\varnothing\rangle=|\beta_{q}|^{2}\cdot N.\label{eq:Fock-N}
\end{equation}
As a pulse-shaped multi-mode state, indeed generally $|N,\{\beta_{k}\}\rangle$
is not the eigenstate of $\hat{\boldsymbol{N}}=\sum\hat{a}^{\dag}_{k}\hat{a}_{k}$
(total photon number operator of the EM field), while its mean photon
number is $\langle\hat{\boldsymbol{N}}\rangle_{N,\{\beta_{k}\}}=N$.
This is because the photon number in the pulse also depends on the
mode distribution $|\beta_{q}|^{2}$ {[}Eq.\,(\ref{eq:Fock-N}){]}.
On the other hand, it is the eigenstate of $\hat{A}^{\dag}\hat{A}=\sum\,|\beta_{k}|^{2}\,\hat{a}^{\dag}_{k}\hat{a}_{k}$,
namely, $\hat{A}^{\dag}\hat{A}\,|N,\{\beta_{k}\}\rangle=N\,|N,\{\beta_{k}\}\rangle$. 

Further, we can write down the displacement operator based on the
collective photon operator $\hat{A}$, that is, 
\begin{equation}
\hat{\mathbf{D}}(\alpha):=e^{\alpha\hat{A}^{\dag}-\alpha^{*}\hat{A}}=\prod_{k}e^{\alpha\beta_{k}\,\hat{a}^{\dag}_{k}-(\alpha\beta_{k})^{*}\,\hat{a}}=\prod_{k}\hat{D}_{k}(\alpha\beta_{k})=e^{-\frac{1}{2}|\alpha|^{2}}\cdot e^{\alpha\hat{A}^{\dag}}\cdot e^{-\alpha^{*}\hat{A}}.
\end{equation}
Then the multi-mode coherent pulse state (\ref{eq:coherent-pulse})
in the main text also can be written as $\big|\alpha,\,\{\beta_{k}\}\big\rangle=\hat{\mathbf{D}}(\alpha)\,|\varnothing\rangle$. 

With these preparations, it can be verified that the above pulse-shaped
Fock state $\big|N,\,\{\beta_{k}\}\big\rangle$ {[}Eq.\,(\ref{eq:Fock-N}){]}
also can be represented by the following pulse-shaped \emph{P}-function
representation, 
\begin{align}
\boldsymbol{\rho}_{|N\rangle} & =\int\text{d}^{2}\alpha\,P_{|N\rangle}(\alpha,\alpha^{*})\,\big|\alpha,\,\{\beta_{k}\}\big\rangle\big\langle\alpha,\,\{\beta_{k}\}\big|=\int\text{d}^{2}\alpha\,\Big[\frac{e^{|\alpha|^{2}}}{N!}\partial^{N}_{\alpha}\partial^{N}_{\alpha^{*}}\delta^{(2)}(\alpha)\Big]\,\big|\alpha,\,\{\beta_{k}\}\big\rangle\big\langle\alpha,\,\{\beta_{k}\}\big|\nonumber \\
 & =\frac{1}{N!}\partial^{N}_{\alpha}\partial^{N}_{\alpha^{*}}\Big[e^{|\alpha|^{2}}\,\hat{\mathbf{D}}(\alpha)\big|\varnothing\big\rangle\big\langle\varnothing\big|\hat{\mathbf{D}}^{\dag}(\alpha)\Big]\,\Big|_{\alpha\rightarrow0}=\frac{1}{N!}\partial^{N}_{\alpha}\partial^{N}_{\alpha^{*}}\Big[e^{\alpha\hat{A}^{\dag}}\big|\varnothing\big\rangle\big\langle\varnothing\big|e^{\alpha^{*}\hat{A}}\Big]\,\Big|_{\alpha\rightarrow0}\nonumber \\
 & =\frac{1}{N!}(\hat{A}^{\dag})^{N}\,|\varnothing\rangle\langle\varnothing|\,\hat{A}^{N}=\big|N,\,\{\beta_{k}\}\big\rangle\big\langle N,\,\{\beta_{k}\}\big|.
\end{align}
Here, $P_{|N\rangle}(\alpha)=(1/N!)\,e^{|\alpha|^{2}}\partial^{N}_{\alpha}\partial^{N}_{\alpha^{*}}\delta^{(2)}(\alpha)$
is the \emph{P}-function for the Fock state widely adopted in quantum
optics \citep{vogel_quantum_2006,walls_quantum_2008,schleich_quantum_2001,agarwal_quantum_2012}.
Clearly, $N=1$ well returns the above single-photon state (\ref{eq:1-photon}).
In this sense, we see the pulse-shaped \emph{P}-function representation
(\ref{eq:P-pulse}) in the main text gives a well description for
a generic quantum light pulse, where the nonclassical properties are
fully enclosed in the single-variable function $P(\alpha,\alpha^{*})$,
and can be generalized as that of any other quantum states. 

\section{The general system dynamics when interacting with an arbitrary quantum
light pulse \label{sec:Master-equation-derivation}}

Here we show the derivation for the master equation in the quasi-classical
$\alpha$-branch. As mentioned in the main text, here we assume the
initial state of the EM field is the multi-mode coherent pulse state
$\hat{\boldsymbol{\rho}}_{\text{\textsc{e}}}(0)=\big|\alpha,\,\{\beta_{k}\}\big\rangle\big\langle\alpha,\,\{\beta_{k}\}\big|$,
and $\hat{\boldsymbol{\rho}}_{\text{\textsc{se}}}(0)=\hat{\rho}_{\text{\textsc{s}}}(0)\otimes\hat{\boldsymbol{\rho}}_{\text{\textsc{e}}}(0)$.
For simplicity, here we consider the quantum system is a TLS, and
its electric dipole operator is $\tilde{\mathbf{d}}(t)=\vec{\wp}\,\tilde{\sigma}^{x}(t)=\vec{\wp}\,\big(|\mathsf{e}\rangle\langle\mathsf{g}|\,e^{i\Omega_{\mathsf{eg}}t}+\mathrm{H.c.}\big)$.
In the interaction picture, the electric field operator is separated
as its mean value and quantum fluctuation $\tilde{\mathbf{E}}(t)=\vec{E}_{\alpha}(t)+\tilde{\mathbf{E}}^{(0)}(t)$,
and the interaction between the TLS and the EM field is described
by {[}see discussions around Eqs.\,(\ref{eq:E(rt)}, \ref{eq:V-alpha})
in the main text{]} 
\begin{align}
\tilde{V}_{\text{\textsc{se}}}(t) & =-\tilde{\mathbf{d}}(t)\cdot\tilde{\mathbf{E}}(t)=\tilde{\sigma}^{x}(t)\cdot\big(\sum_{k}g_{k}\,\hat{a}_{k}\,e^{-i\omega_{k}t}+\mathrm{H.c.}\big):=\tilde{V}_{\alpha}(t)+\tilde{V}^{(0)}_{\text{\textsc{se}}}(t),\nonumber \\
\tilde{V}_{\alpha}(t) & =\tilde{\sigma}^{x}(t)\cdot\big(\sum_{k}g_{k}\,\alpha\beta_{k}\,e^{-i\omega_{k}t}+\mathrm{c.c.}\big):=\tilde{\sigma}^{x}(t)\cdot\mathscr{E}_{\alpha}(t),\\
\tilde{V}^{(0)}_{\text{\textsc{se}}}(t) & =\tilde{\sigma}^{x}(t)\cdot\big(\sum_{k}g_{k}\,\delta\hat{a}_{k}\,e^{-i\omega_{k}t}+\mathrm{H.c.}\big)\simeq\sum_{k}g_{k}\,\hat{\sigma}^{+}\,\delta\hat{a}_{k}\,e^{i(\Omega_{\mathsf{eg}}-\omega_{k})t}+\mathrm{H.c.}:=\tilde{V}^{(+)}_{\text{\textsc{se}}}(t)+\tilde{V}^{(-)}_{\text{\textsc{se}}}(t).\nonumber 
\end{align}

Here the interaction $\tilde{V}_{\alpha}(t)$ only contains the operators
for the TLS, thus can be regarded as the quasi-classical driving interaction.
The dynamics of the TLS can be obtained by taking the integral iteration
of the von Neumann equation \citep{breuer_theory_2002,li_non-markovianity_2016},
\begin{align}
\partial_{t}\tilde{\boldsymbol{\rho}}_{\text{\textsc{se}}}(t) & =\frac{i}{\hbar}[\tilde{\boldsymbol{\rho}}_{\text{\textsc{se}}}(t),\,\tilde{V}_{\text{\textsc{se}}}(t)]=\frac{i}{\hbar}[\tilde{\boldsymbol{\rho}}_{\text{\textsc{se}}}(t),\,\tilde{V}_{\alpha}(t)]+\frac{i}{\hbar}[\tilde{\boldsymbol{\rho}}_{\text{\textsc{se}}}(t),\,\tilde{V}^{(0)}_{\text{\textsc{se}}}(t)],\nonumber \\
\tilde{\boldsymbol{\rho}}_{\text{\textsc{se}}}(t) & =\tilde{\boldsymbol{\rho}}_{\text{\textsc{se}}}(0)+\frac{i}{\hbar}\int^{t}_{0}\text{d}s\,\big[\,\tilde{\boldsymbol{\rho}}_{\text{\textsc{se}}}(s),\,\tilde{V}_{\alpha}(s)+\tilde{V}^{(0)}_{\text{\textsc{se}}}(s)\,\big].
\end{align}
Now we put the integral form of $\tilde{\boldsymbol{\rho}}_{\text{\textsc{se}}}(t)$
back into the second commutator bracket in the first line, and that
gives \begin{subequations}
\begin{align}
\partial_{t}\tilde{\boldsymbol{\rho}}_{\text{\textsc{se}}}(t) & =\frac{i}{\hbar}[\tilde{\boldsymbol{\rho}}_{\text{\textsc{se}}}(t),\,\tilde{V}_{\alpha}(t)]+\frac{i}{\hbar}[\tilde{\boldsymbol{\rho}}_{\text{\textsc{se}}}(0),\,\tilde{V}^{(0)}_{\text{\textsc{se}}}(t)]-\frac{1}{\hbar^{2}}\int^{t}_{0}\text{d}s\,\Big[\,\big[\,\tilde{\boldsymbol{\rho}}_{\text{\textsc{se}}}(s),\,\tilde{V}_{\alpha}(s)+\tilde{V}^{(0)}_{\text{\textsc{se}}}(s)\,\big],\,\tilde{V}^{(0)}_{\text{\textsc{se}}}(t)\,\Big],\label{eq:ME-se}\\
\partial_{t}\tilde{\mathbf{\rho}}_{\text{\textsc{s}}}(t) & \simeq\frac{i}{\hbar}[\tilde{\mathbf{\rho}}_{\text{\textsc{s}}}(t),\,\tilde{V}_{\alpha}(t)]-\frac{1}{\hbar^{2}}\int^{t}_{0}\text{d}s\,\mathrm{tr}_{\text{\textsc{e}}}\,\Big[\,\big[\,\tilde{\mathbf{\rho}}_{\text{\textsc{s}}}(t)\otimes\tilde{\boldsymbol{\rho}}_{\text{\textsc{e}}}(0),\,\tilde{V}^{(0)}_{\text{\textsc{se}}}(s)\,\big],\,\tilde{V}^{(0)}_{\text{\textsc{se}}}(t)\,\Big].\label{eq:ME-s}
\end{align}
\end{subequations} The second line is obtained by tracing out the
EM field, which makes the second bracket in Eq.\,(\ref{eq:ME-se})
directly vanishes. And here we apply the Born-Markovian approximation
$\tilde{\boldsymbol{\rho}}_{\text{\textsc{se}}}(s)\mapsto\tilde{\mathbf{\rho}}_{\text{\textsc{s}}}(t)\otimes\tilde{\boldsymbol{\rho}}_{\text{\textsc{e}}}(0)$
\citep{breuer_theory_2002,li_non-markovianity_2016}. The integral
term in Eq.\,(\ref{eq:ME-s}) contains an iterated commutator bracket,
and here we show the calculation for one of the expanded terms: 
\begin{align}
\int^{t}_{0}\text{d}s\,\mathrm{tr}_{\text{\textsc{e}}}\big[\,\tilde{\mathbf{\rho}}_{\text{\textsc{s}}}(t)\otimes\tilde{\boldsymbol{\rho}}_{\text{\textsc{e}}}(0)\,\tilde{V}^{(+)}_{\text{\textsc{se}}}(s)\,\tilde{V}^{(-)}_{\text{\textsc{se}}}(t)\,\big] & =\tilde{\mathbf{\rho}}_{\text{\textsc{s}}}(t)\hat{\sigma}^{+}\hat{\sigma}^{-}\cdot\sum_{k,q}g_{k}g^{*}_{q}\int^{t}_{0}\text{d}s\,\mathrm{tr}_{\text{\textsc{e}}}\big[\hat{\boldsymbol{\rho}}_{\text{\textsc{e}}}(0)\,\delta\hat{a}_{k}\,\delta\hat{a}^{\dag}_{q}\big]\,e^{i(\Omega_{\mathsf{eg}}-\omega_{k})s-i(\Omega_{\mathsf{eg}}-\omega_{q})t}\nonumber \\
 & =\tilde{\mathbf{\rho}}_{\text{\textsc{s}}}(t)\hat{\sigma}^{+}\hat{\sigma}^{-}\cdot\sum_{k}|g_{k}|^{2}\,\langle\delta\hat{a}_{k}\,\delta\hat{a}^{\dag}_{k}\rangle_{\alpha,\{\beta_{k}\}}\int^{t}_{0}\text{d}s\,e^{i(\Omega_{\mathsf{eg}}-\omega_{k})(s-t)}.
\end{align}
As discussed around Eqs.\,(\ref{eq:E(rt)}, \ref{eq:V-alpha}), for
the field state $\hat{\boldsymbol{\rho}}_{\text{\textsc{e}}}(0)=\big|\alpha,\,\{\beta_{k}\}\big\rangle\big\langle\alpha,\,\{\beta_{k}\}\big|$,
its fluctuation satisfies $\langle\delta\hat{a}_{k}\,\delta\hat{a}^{\dag}_{k}\rangle_{\alpha,\{\beta_{k}\}}=\langle\varnothing|\,\hat{a}_{k}\hat{a}^{\dag}_{k}\,|\varnothing\rangle$,
which is the same with the situation when considering the spontaneous
emission of the TLS in the vacuum field without any driving light.
Therefore, the integral term in Eq.\,(\ref{eq:ME-s}) would just
give the standard GKSL (Lindblad) terms that have the same form with
the Born-Markovian master equation when considering the atom dissipations
without any driving light. Thus, the TLS dynamics {[}Eq.\,(\ref{eq:ME-s}){]}
finally gives the master equation (\ref{eq:TLS-ME}) in the main text.

In this sense, when the initial state of the EM field is the multi-mode
coherent pulse state, the TLS dynamics can be described by the master
equation with the quasi-classical driving term, which well returns
the description widely adopted in literatures. 

\section{A two-level system driven by a classical coherent pulse: the Bloch
equations \label{sec:classical}}

Here we study the time-dependent evolution of a TLS, which is interacting
with a classical coherent pulse. Here we consider the input light
is described by the wave packet (\ref{eq:E(xt)}) with the exponential
shape in the time domain, and the TLS gap is resonant with the central
frequency of the light pulse $\Omega_{\mathsf{eg}}=\omega_{\text{d}}$.
As mentioned in Sec.\,\ref{subsec:Bloch-0-phase}, denoting $u\equiv\langle\hat{\sigma}^{x}\rangle$,
$v\equiv\langle\hat{\sigma}^{y}\rangle$, $w\equiv\langle\hat{\sigma}^{z}\rangle$,
when the driving phase is zero ($\mathcal{E}_{\alpha}\equiv\alpha\,\eta_{0}/\hbar>0$),
their evolutions follow the Bloch equations (\ref{eq:bloch-u}-\ref{eq:bloch-w}). 

The equation (\ref{eq:bloch-u}) for $u(t)$ directly gives $u(t)=u(0)\,e^{-\frac{1}{2}\gamma t}$.
Now we focus on the evolution of $v(t)$ and $w(t)$. For these two
equations, we rescale the time $t$ by a variable transformation $\zeta=e^{-\frac{1}{2}\kappa t}\in(0,1]$,
$\frac{\text{d}}{\text{d}t}=-\frac{1}{2}\kappa\zeta\frac{\text{d}}{\text{d}\zeta}$,
and they become \citep{allen_optical_1987,zlatanov_exact_2015} 
\begin{align}
\frac{\text{d}v}{\text{d}\zeta} & =-\frac{2\mathcal{E}_{\alpha}}{\kappa}\,w+\frac{\gamma}{\kappa}\,\frac{v}{\zeta},\nonumber \\
\frac{\text{d}w}{\text{d}\zeta} & =\frac{2\mathcal{E}_{\alpha}}{\kappa}\,v+\frac{2\gamma}{\kappa}\,\frac{w+1}{\zeta}.
\end{align}
 Taking $v[w,w']$ from the second equation, and substituting it back
to the first one, that gives (denoting $\tilde{\gamma}\equiv\gamma/\kappa$,
$\tilde{\mathcal{E}}\equiv\mathcal{E}_{\alpha}/\kappa$)
\begin{equation}
\zeta^{2}\,w''-3\tilde{\gamma}\,\zeta\,w'+\big[4\tilde{\mathcal{E}}^{2}\,\zeta^{2}+2\tilde{\gamma}(\tilde{\gamma}+1)\big]\,w=-2\tilde{\gamma}(\tilde{\gamma}+1).
\end{equation}

To make a more convenient discussion about the boundary conditions
of time, hereafter we study $\bar{w}(\zeta):=w(\zeta)+1$ instead
of $w(\zeta)=\langle\hat{\sigma}^{z}\rangle$. Initially at $t=0\,(\zeta=1)$,
the atom is in the ground state; when $t\rightarrow\infty\,(\zeta=0)$,
the atom would decay back to the ground state after the excitation.
Thus, the boundary condition is $\bar{w}\big|_{\zeta=0}=\bar{w}\big|_{\zeta=1}=0$,
and the equation of $\bar{w}(\zeta)$ is 
\begin{equation}
\zeta^{2}\,\bar{w}''-3\tilde{\gamma}\,\zeta\,\bar{w}'+\big[4\tilde{\mathcal{E}}^{2}\,\zeta^{2}+2\tilde{\gamma}(\tilde{\gamma}+1)\big]\,\bar{w}=4\tilde{\mathcal{E}}^{2}\,\zeta^{2}.
\end{equation}
Now we further apply the Fr\"obenius method by setting $\bar{w}(\zeta):=\zeta^{p}\,y(\zeta)$
with $p=\frac{1}{2}+\frac{3}{2}\tilde{\gamma}$, and that further
gives an inhomogeneous Bessel equation ($\nu:=\tfrac{1}{2}|\tilde{\gamma}-1|$)
\begin{equation}
y''+\frac{1}{\zeta}\,y'+(4\tilde{\mathcal{E}}^{2}-\frac{\nu^{2}}{\zeta^{2}})\,y=4\tilde{\mathcal{E}}^{2}\,\zeta^{-p}.\label{eq:bessel}
\end{equation}

For the homogeneous part of Eq.\,(\ref{eq:bessel}), the solutions
are the Bessel functions $J_{\nu}(2\tilde{\mathcal{E}}\,\zeta)$ and
$Y_{\nu}(2\tilde{\mathcal{E}}\,\zeta)$. Then the full solution of
$\bar{w}(\zeta)$ can be obtained with the help the Wronskian determinant
method \citep{weber_mathematical_2008}, that is, 
\begin{equation}
\bar{w}(\zeta)=2\pi\,\tilde{\mathcal{E}}^{2}\,\zeta^{p}\,\Big[Y_{\nu}(2\tilde{\mathcal{E}}\,\zeta)\int^{\zeta}_{1}J_{\nu}(2\tilde{\mathcal{E}}\,\eta)\,\eta^{1-p}\,\text{d}\eta-J_{\nu}(2\tilde{\mathcal{E}}\,\zeta)\int^{\zeta}_{1}Y_{\nu}(2\tilde{\mathcal{E}}\,\eta)\,\eta^{1-p}\,\text{d}\eta\Big].\label{eq:wb(s)}
\end{equation}
Here the integral lower bounds are particularly set in such a way
that the boundary conditions $\bar{w}\big|_{\zeta=0}=\bar{w}\big|_{\zeta=1}=0$
are directly satisfied. Therefore, this is just the analytical result
for $\bar{w}(\zeta)$ in the above Bloch equation (\ref{eq:bloch-w}).
By substituting this result $\bar{w}(\zeta)$ back into the Bloch
equation (\ref{eq:bloch-v}) for $v(t)$, the analytical solution
of $v(t)$ also can be obtained immediately.

Further, by utilizing series form of the Bessel functions, the above
integral function in Eq.\,(\ref{eq:wb(s)}) can be expanded as 
\begin{align}
Y_{\nu}(2\tilde{\mathcal{E}}\,\zeta) & J_{\nu}(2\tilde{\mathcal{E}}\,\eta)-J_{\nu}(2\tilde{\mathcal{E}}\,\zeta)Y_{\nu}(2\tilde{\mathcal{E}}\,\eta)=\sum^{\infty}_{K=0}C_{K}(\zeta,\eta)\cdot\tilde{\mathcal{E}}^{2K}\nonumber \\
C_{K}(\zeta,\eta) & :=\frac{(-1)^{K}}{\sin(\nu\pi)}\sum^{K}_{q=0}\frac{\zeta^{2K}(\eta/\zeta)^{2q+\nu}-\eta^{2K}(\zeta/\eta)^{2q+\nu}}{q!(K-q)!\Gamma(q+\nu+1)\Gamma(K+1-q-\nu)}.
\end{align}
 Substituting these power terms with $\eta^{a}$ back into the above
integral in Eq.\,(\ref{eq:wb(s)}), we obtain the expansion form
$\bar{w}(\zeta)=\sum^{\infty}_{K=1}\,\mathscr{C}_{K}(\zeta)\,\tilde{\mathcal{E}}^{2K}$
presented in the main text {[}Eq.\,(\ref{eq:wb-expand}){]}, where
we used the relation $\Gamma(1+\nu)\Gamma(1-\nu)=\pi/\sin(\nu\pi)$. 

Particularly, the coefficient $\mathscr{C}_{K=1}(\zeta)$ is 
\begin{equation}
\begin{split}\mathscr{C}_{K=1}(\zeta)=\frac{2}{\nu}\,\big(\frac{\zeta^{2}-\zeta^{p+\nu}}{2-p-\nu}-\frac{\zeta^{2}-\zeta^{p-\nu}}{2-p+\nu}\big) & =\frac{2(\zeta-\zeta^{\tilde{\gamma}})^{2}}{(1-\tilde{\gamma})^{2}}=2\kappa^{2}\cdot\frac{(e^{-\frac{1}{2}\kappa t}-e^{-\frac{1}{2}\gamma t})^{2}}{(\kappa-\gamma)^{2}},\\
\Rightarrow\qquad\lim_{\gamma\rightarrow\kappa}\mathscr{C}_{K=1}(\zeta) & =2(\zeta\ln\zeta)^{2}=\frac{1}{2}\kappa^{2}t^{2}\,e^{-\kappa t}.
\end{split}
\label{eq:D_1}
\end{equation}
This result is used in Eqs.\,(\ref{eq:w1p}, \ref{eq:Ne(t)-1p})
when studying the TLS dynamics driven by a nonclassical single-photon
pulse (Sec.\,\ref{subsec:Benchmark-1p}). 

\section{A two-level system driven by a single-photon pulse: the fully quantum
treatment \label{sec:quantum}}

Now we study the dynamics of a TLS interacting with a single photon
pulse by adopting the fully quantum approach \citep{wang_efficient_2011,liao_spectrum_2012,liao_single-photon_2015,stobinska_perfect_2009}.
Here we need the Hamiltonian of the TLS and the fully quantized 1D
EM field, which is described by (after RWA) 
\begin{equation}
\hat{\mathcal{H}}=\frac{1}{2}\hbar\Omega_{\mathsf{eg}}\,\hat{\sigma}^{z}+\sum_{k}\,\big(g_{k}\,\hat{\sigma}^{+}\hat{a}_{k}+g^{*}_{k}\,\hat{\sigma}^{-}\hat{a}^{\dag}_{k})+\sum_{k}\hbar\omega_{k}\,\hat{a}^{\dag}_{k}\hat{a}_{k},\qquad g_{k}=-i\,\wp\,\sqrt{\frac{\hbar\omega_{k}}{2\epsilon_{0}V}},\label{eq:H-A-field}
\end{equation}
 For this system, the total excitation operator $\hat{N}:=|\mathsf{e}\rangle\langle\mathsf{e}|+\sum\hat{a}^{\dag}_{k}\hat{a}_{k}$
is conserved ( $[\hat{N},\,\hat{\mathcal{H}}]=0$ ). Thus, the evolution
process is constrained in a subspace determined by the total excitation
number, which can be well solved. 

We assume initially the atom is in the ground state $|\mathsf{g}\rangle$,
and the EM field is in the single photon pulse state $|\Psi_{1\mathrm{p}}\rangle=\sum_{k}\,\beta_{k}\,\hat{a}^{\dag}_{k}|\varnothing\rangle$
{[}Eq.\,(\ref{eq:1-photon}){]}, where $\{\beta_{k}\}$ takes the
Lorentzian distribution (\ref{eq:beta-k}). The total excitation conservation
guarantees the full atom-field state always can be written as the
following form, 
\begin{equation}
|\Psi_{t}\rangle=A_{\mathsf{e}}(t)\,e^{-\frac{i}{2}\Omega_{\mathsf{eg}}t}\,|\mathsf{e}\rangle\otimes|\varnothing\rangle+\sum_{k}\beta_{k}(t)\,e^{-i\omega_{k}t+\frac{i}{2}\Omega_{\mathsf{eg}}t}\,|\mathsf{g}\rangle\otimes|1_{k}\rangle.
\end{equation}
Here $A_{\mathsf{e}}(0)=0$, and the state $|\Psi_{t}\rangle$ follows
the unitary evolution governed by the Hamiltonian (\ref{eq:H-A-field}),
which gives 
\begin{equation}
\begin{split}\partial_{t}A_{\mathsf{e}} & =-\frac{i}{\hbar}\sum_{k}g_{k}\,e^{-i(\omega_{k}-\Omega_{\mathsf{eg}})t}\,\beta_{k}(t),\\
\partial_{t}\beta_{k} & =-\frac{i}{\hbar}\,g^{*}_{k}\,e^{i(\omega_{k}-\Omega_{\mathsf{eg}})t}\,A_{\mathsf{e}}(t).
\end{split}
\end{equation}
 These two equations further give 
\begin{align}
\beta_{k}(t) & =\beta_{k}(0)-\frac{i}{\hbar}\,g^{*}_{k}\int^{t}_{0}\text{d}s\,e^{i(\omega_{k}-\Omega_{\mathsf{eg}})s}\,A_{\mathsf{e}}(s),\nonumber \\
\partial_{t}A_{\mathsf{e}} & =-\frac{1}{\hbar^{2}}\int^{t}_{0}\text{d}s\,\Big[\sum_{k}|g_{k}|^{2}e^{i(\omega_{k}-\Omega_{\mathsf{eg}})s}\Big]\,A_{\mathsf{e}}(s)-\frac{i}{\hbar}\sum_{k}g_{k}\,\beta_{k}(0)\,e^{-i(\omega_{k}-\Omega_{\mathsf{eg}})t}\nonumber \\
 & \simeq-\frac{\gamma}{2}\,A_{\mathsf{e}}+\frac{i\eta_{0}}{2\hbar}\,e^{i(\Omega_{\mathsf{eg}}-\omega_{\text{d}})t-\frac{1}{2}\kappa t},\qquad\eta_{0}\equiv\wp\sqrt{\frac{2\hbar\omega_{\text{d}}}{\epsilon_{0}(\mathsf{S}c/\kappa)}}\equiv\wp\,\mathtt{E}_{0},\label{eq:dt_Ae(t)}
\end{align}
where the last line is obtained by the following calculations, 
\begin{align}
\frac{1}{\hbar^{2}}\int^{t}_{0}\text{d}s\,\sum_{k}|g_{k}|^{2}e^{i(\omega_{k}-\Omega_{\mathsf{eg}})s} & =\frac{1}{\hbar^{2}}\cdot\frac{L}{2\pi}\int^{\infty}_{-\infty}\text{d}k\ \frac{\wp^{2}\,\hbar\omega_{k}}{2\epsilon_{0}V}\ \big[\pi\delta(\omega_{k}-\Omega_{\mathsf{eg}})+i\mathbf{P}\frac{1}{\omega_{k}-\Omega_{\mathsf{eg}}}\big]\nonumber \\
 & \simeq\frac{1}{2}\frac{\wp^{2}\,\Omega_{\mathsf{eg}}}{\hbar\epsilon_{0}\mathsf{S}c}:=\frac{1}{2}\gamma,\qquad\gamma:=\frac{\wp^{2}\,\Omega_{\mathsf{eg}}}{\hbar\epsilon_{0}\mathsf{S}c},\label{eq:decay-gamma}
\end{align}
\begin{align}
\sum_{k}g_{k}\,\beta_{k}(0)\,e^{-i\omega_{k}t} & =\frac{L}{2\pi}\int^{\infty}_{-\infty}\text{d}k\:\Big(-i\wp\sqrt{\frac{\hbar\omega_{k}}{2\epsilon_{0}V}}\,\Big)\cdot\sqrt{\frac{c\kappa}{L}}\frac{1}{ck-\omega_{\text{d}}+i\kappa/2}\,e^{-ickt}\nonumber \\
 & =-\frac{1}{2}\,\wp\,\sqrt{\frac{2\hbar\omega_{\text{d}}}{\epsilon_{0}(\mathsf{S}c/\kappa)}}\:\Theta(t)e^{-i\omega_{\text{d}}t-\frac{1}{2}\kappa t}\equiv-\frac{\eta_{0}}{2}\:e^{-i\omega_{\text{d}}t-\frac{1}{2}\kappa t}.\label{eq:force}
\end{align}

Here we consider the TLS gap and the central frequency of the pulse
are resonant $\Omega_{\mathsf{eg}}=\omega_{\text{d}}$. Eq.\,(\ref{eq:decay-gamma})
gives the the spontaneous emission rate $\gamma$ of the TLS in this
1D EM field, where the photons can be emitted to both left and right
directions (such a bi-direction consideration $\omega_{k}=c|k|$ contributes
a factor $2$ for the density of states in this integral), and the
principle integral is omitted. The term Eq.\,(\ref{eq:force}) provides
a driving force for $\partial_{t}A_{\mathsf{e}}$ {[}Eq.\,(\ref{eq:dt_Ae(t)}){]}
from the light pulse, where $\eta_{0}\equiv\wp\,\mathtt{E}_{0}$ is
the effective coupling strength between the TLS and a single photon
in the pulse. In the above derivations, it is worth noting that, the
TLS decay rate $\gamma$, the pulse linewidth $\kappa$, and the single-photon
coupling strength $\eta_{0}$ have an important relation $\eta_{0}=\hbar\sqrt{2\gamma\kappa}$.
Finally, the solution of Eq.\,(\ref{eq:dt_Ae(t)}) gives 
\begin{align}
A_{\mathsf{e}}(t) & =A_{\mathsf{e}}(0)\,e^{-\frac{1}{2}\gamma t}+\frac{i\eta_{0}}{2\hbar}\int^{t}_{0}\text{d}s\,e^{-\frac{1}{2}\gamma(t-s)}\cdot e^{-\frac{1}{2}\kappa s}=\frac{i\sqrt{2\gamma\kappa}}{\kappa-\gamma}(e^{-\frac{1}{2}\gamma t}-e^{-\frac{1}{2}\kappa t}),\nonumber \\
\bar{\text{\textsc{n}}}_{\mathsf{e}}(t) & =\big|A_{\mathsf{e}}(t)\big|^{2}=2\gamma\kappa\cdot\frac{(e^{-\frac{1}{2}\gamma t}-e^{-\frac{1}{2}\kappa t})^{2}}{(\gamma-\kappa)^{2}}.\label{eq:P_e(t)}
\end{align}

Particularly, in the limit $\gamma=\kappa$, the above result further
gives $\bar{\text{\textsc{n}}}_{\mathsf{e}}(t)=\frac{1}{2}\gamma^{2}t^{2}\,e^{-\gamma t}$.
In this case, when $\gamma t=2$, the TLS reaches its maximum excitation
probability $\bar{\text{\textsc{n}}}^{\text{max}}_{\mathsf{e}}=2/e^{2}\simeq0.27$.
This result is well consistent with previous studies in literature.
Now we see the approach of the \emph{P}-function average gives the
same time-dependent results as here {[}see discussions around Eq.\,(\ref{eq:Ne(t)-1p}){]}.

\end{widetext}

\bibliography{Refs,liao}

\end{document}